\documentclass[sigplan,screen,nonacm]{acmart}

\usepackage{balance}  % for  \balance command ON LAST PAGE  (only there!)

\usepackage{amsmath,amssymb,amsfonts}
\usepackage{algorithmic}
\usepackage{graphicx}
\usepackage{textcomp}
\usepackage{xcolor}
\usepackage{booktabs} % For formal tables
\usepackage{hyperref}
\usepackage{amsmath}
\usepackage{enumitem}
\usepackage{relsize}
\usepackage{subcaption}

\newtheorem{definition}{Definition}[section]

\begin{document}
	
	% ****************** TITLE ****************************************
	
	\title{qwLSH: Cache-conscious Indexing for Processing Similarity Search Query Workloads in High-Dimensional Spaces}
	
	\titlenote{This work is supported by NSF \#1633330}
		\author{Omid Jafari}
		\email{ojafari@nmsu.edu}
		\affiliation{%
			\institution{New Mexico State University}
			\streetaddress{Computer Science Department}
			\city{Las Cruces}
			\state{New Mexico}
			\postcode{88003}
		}
	
		\author{John Ossorgin}
		\email{osso09@nmsu.edu}
		\affiliation{%
			\institution{New Mexico State University}
			\streetaddress{Computer Science Department}
			\city{Las Cruces}
			\state{New Mexico}
			\postcode{88003}
		}
		
		\author{Parth Nagarkar}
		\email{nagarkar@nmsu.edu}
		\affiliation{%
			\institution{New Mexico State University}
			\streetaddress{Computer Science Department}
			\city{Las Cruces}
			\state{New Mexico}
			\postcode{88003}
		}
	
	\begin{abstract}
		Similarity search queries in high-dimensional spaces are an important type of queries in many domains such
		as image processing, machine learning, etc. Since exact similarity search indexing techniques suffer from the well-known \textit{curse of dimensionality} in high-dimensional spaces, approximate search techniques are often utilized instead. Locality Sensitive Hashing (LSH) has been shown to be an effective approximate search method for solving similarity search queries in high-dimensional spaces. Often times, queries in real-world settings arrive as part of a query workload. LSH and its variants are particularly designed to solve single queries effectively. They suffer from one major drawback while executing query workloads: they do not take into consideration important data characteristics for effective cache utilization while designing the index structures. In this paper, we present \textit{qwLSH}, an index structure for efficiently processing similarity search query workloads in high-dimensional spaces. We intelligently divide a given cache during processing of a query workload by using novel cost models. Experimental results show that, given a query workload, \textit{qwLSH} is able to perform faster than existing techniques due to its unique cost models and strategies.
		 %We further present different caching strategies for efficiently processing similarity search query workloads. We evaluate our proposed unique design and cost models of \textit{qwLSH} on real datasets against state-of-the-art LSH-based techniques. 
	\end{abstract}

	%
	% Keywords. The author(s) should pick words that accurately describe the work being
	% presented. Separate the keywords with commas.

\maketitle

\section{Introduction}
The similarity search problem in high-dimensional spaces is a well-known problem with wide-ranging applications in domains such as information retrieval, artificial intelligence, machine learning, etc. Exact tree-based spatial indexing techniques, such as KD-tree, R-tree, SR-tree, etc., are effective for improving searching in low-dimensional spaces, but as the dimensions increase ($\sim>$10), they all suffer from the popular \textit{curse of dimensionality} problem (where they are often outperformed even by a brute force linear scan) \cite{Datar:2004:LHS:997817.997857}. One approach to addressing this \textit{curse of dimensionality} problem is to search for \textit{approximate} solutions instead of exact solutions. In many applications, where 100\% accuracy is unnecessary, \textit{good enough} results are acceptable. Approximate solutions sacrifice accuracy for a much faster performance. Formally, the goal of the approximate version of the similarity search problem, also called \textit{c-approximate Nearest Neighbor search}, is to return the  objects that are within the distance $c \times R$ from the query object, where $c>1$ is an approximate ratio and $R$ is the distance from the query to the true nearest neighbor. 

%\vspace*{-0.02in}	

\subsection{Locality Sensitive Hashing}
Locality Sensitive Hashing, first proposed in \cite{Indyk:1998:ANN:276698.276876}, is one of the most popular solutions for approximate searching in high-dimensional spaces. The purpose of Locality Sensitive Hashing (LSH) is to map high-dimensional data to lower dimensions while preserving the distances in the original space. The lower dimensional space is generated through a series of \textit{random projections}. In this lower dimensional space, data objects are mapped to individual buckets based on a hash function, with the intuition that nearby data points in the original space are mapped to the same hash buckets in the lower dimensional space with a higher probability than mapping dissimilar or far away points to the same buckets. This may lead to \textit{misses} and \textit{false positives}. Given a distance metric and a
corresponding LSH family (detailed in Section \ref{sec:prelim}), LSH data structures control their precision and recall by using multiple
independently chosen hash functions organized into several hash layers.
%: intuitively, conjunctively combined hashes at each
%layer reduce false positives, whereas disjunctively combined layers of hash functions help avoid misses – often, at the
%expense of identifying candidate data elements that needs to be eliminated during post-processing.
Since the original work \cite{Indyk:1998:ANN:276698.276876} was proposed, there has been considerable amount of research done on improving Locality Sensitive Hashing \cite{Datar:2004:LHS:997817.997857, Bawa:2005:LFS:1060745.1060840, Liu:2014:SEI:2732939.2732947, Gao:2015:SHC:2783258.2783284, Gan:2012:LHS:2213836.2213898, Sun:2014:SSC:2735461.2735462, Tao:2010:EAN:1806907.1806912, Zheng:2016:LAN:2882903.2882930, Huang:2015:QLH:2850469.2850470}.

\begin{figure}
	\centering
	\includegraphics[width=0.5\linewidth]{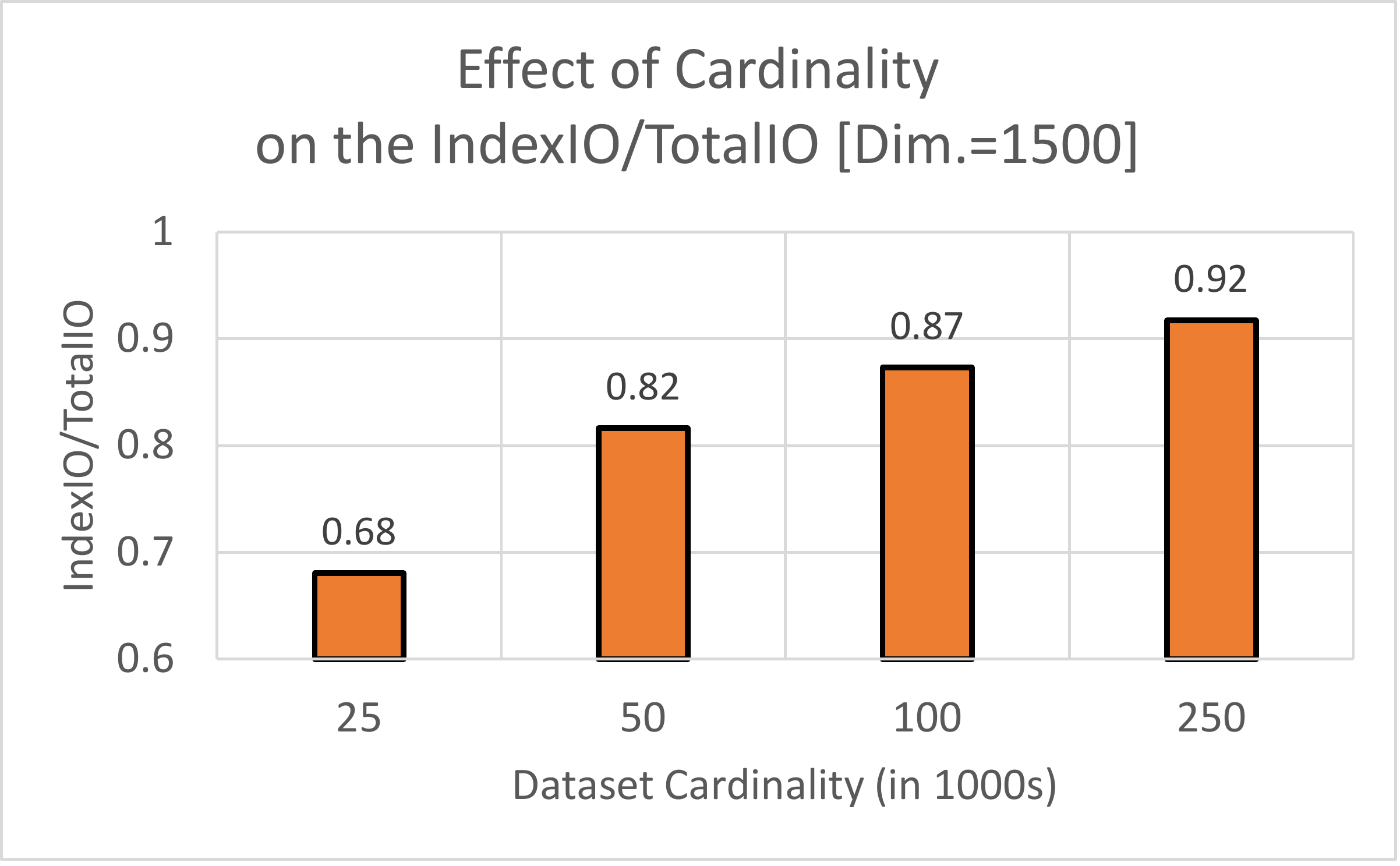}
	\caption{Effect of Cardinality over the ratio IndexIO/TotalIO. We create versions of the Deepsat dataset (see Section \ref{sec:eval} for more details) with different cardinalities but same dim. [=1500] for 250 top-50 queries using QALSH alg. \cite{Gao:2015:SHC:2783258.2783284}.}
	\label{fig:motivationCard}
\end{figure}

\begin{figure}
	\centering
	\includegraphics[width=0.5\linewidth]{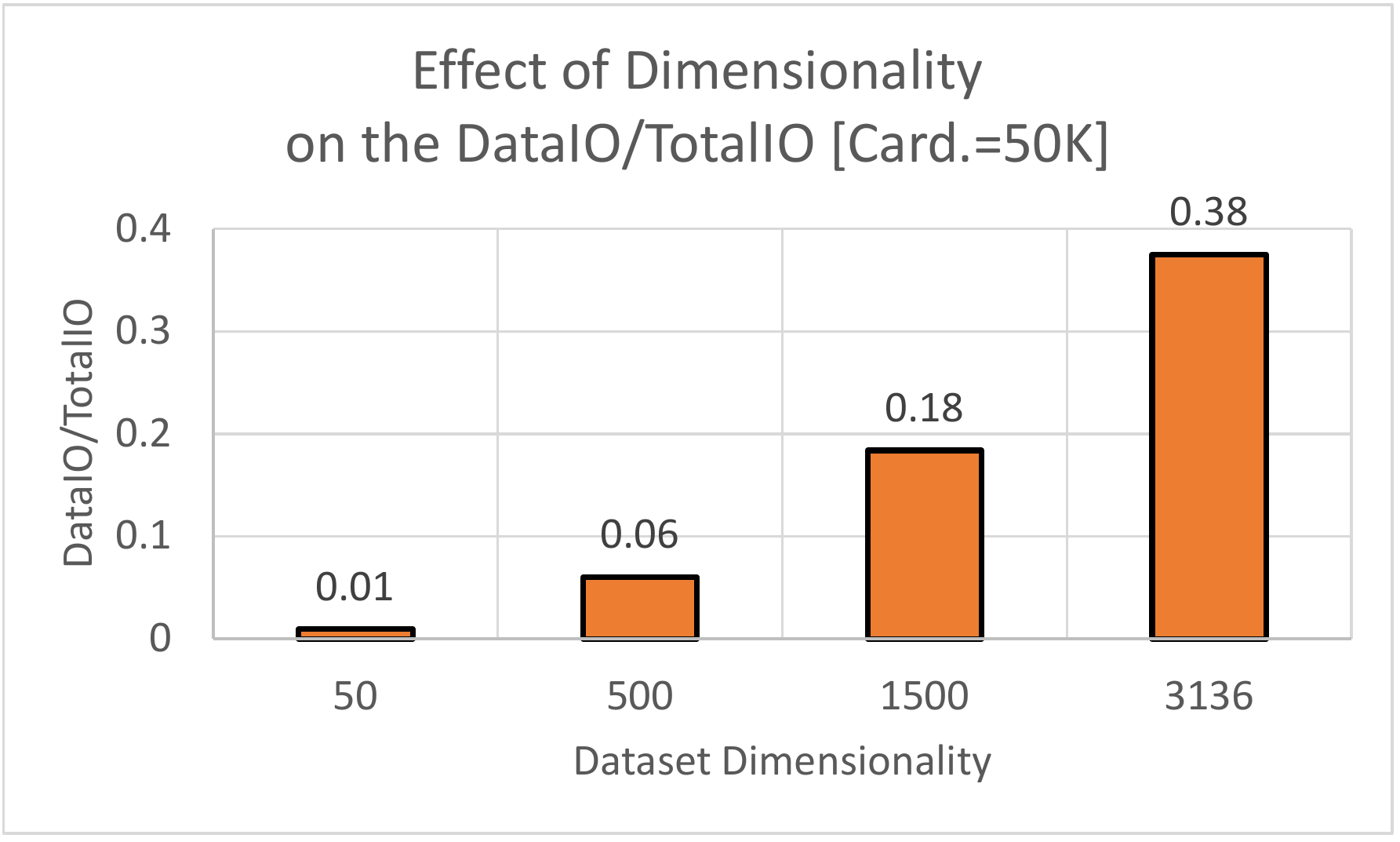}
	\caption{Effect of Dimensionality over DataIO/TotalIO over the Deepsat dataset with different dimensionalities but same card. [=50K] for 250 top-50 queries using QALSH alg. \cite{Gao:2015:SHC:2783258.2783284}.}
	\label{fig:motivationDim}
\end{figure}

\subsection{Motivation}
\label{sec:motiv}
Often times, in real-world settings, queries arrive as part of a query workload. Several research works have shown the benefits of designing index structures particularly for efficient handling of query workloads \cite{Curino:2010:SWA:1920841.1920853, Aly:2015:AAQ:2831360.2831361, Tzoumas:2009:WIC:1687627.1687761, DBLP:journals/pvldb/NagarkarCB15, DBLP:conf/edbt/NagarkarC14, Quamar:2013:SSW:2452376.2452427}. Most of these works focus on partitioning the data space such that regions that are queried with high frequency are partitioned with more granularity. e.g., in 2D geographical spaces, points of interest in downtown areas are queried with much more frequency than places in the suburbs. Similarly, consider the following two scenarios for high-dimensional data: (1) Genomic DNA data, often represented as high-dimensional vectors, require similarity search queries to find similar genomes given a query genome \cite{Buhler2001, Berlin2015, Rasheed2013}. Certain common genome sequences are often queried more times than the rest \cite{Rasheed2013}. (2) Similarly, earthquake detection or satellite image data, both high-dimensional data \cite{Yoon2015, Buaba2010}, often query certain similar regions of the space that are of more interest. These queries can be viewed as part of a query workload. 
While LSH has been shown to be effective in processing single queries in high-dimensional spaces, it is not particularly designed to handle query workloads efficiently. 

Assuming that the index and the data are stored on an external storage, Locality Sensitive Hashing and its variants have two main IO operations: (1) accessing the index (denoted by \textit{IndexIO}) in order to find the candidates, and (2) once the candidates are found, the candidate points need to be brought from the external storage (denoted by \textit{DataIO}) into the main memory to remove false positives. We observe that these two operations, however, have different costs: 
\begin{itemize} [leftmargin=*]
	\item the \textit{IndexIO} cost is dependent on the size of the index, which is in turn dependent on the size of the dataset (also referred to as \textit{cardinality} of the dataset), and
	\item the \textit{dataIO} cost is dependent on the size of each data point, i.e. the number of dimensions of the dataset, and the desired number of objects in the result (top-$k$). 
\end{itemize}

Figures \ref{fig:motivationCard} and \ref{fig:motivationDim} show our observation. In Figure \ref{fig:motivationCard}, it can be seen that as the cardinality of the dataset increases (when the dimensionality is fixed), the ratio of \textit{IndexIO}/\textit{TotalIO} (where \textit{TotalIO = IndexIO + DataIO}) increases, because the cost of IndexIO increases. Similarly, in Figure \ref{fig:motivationDim}, it can be seen that as the dimensionality increases, the ratio of \textit{DataIO}/\textit{TotalIO} increases, because the cost of DataIO increases. By using this important observation, the efficiency of an LSH algorithm can be further improved by intelligently utilizing the cache for a given query workload. In this paper, we propose \textit{qwLSH}, an index structure that improves the performance of query workloads in high-dimensional spaces by improving upon the cache utilization of the system. 

\subsection{Contributions of this paper}
Existing LSH-based index structures are designed to efficiently handle single queries. Given a query workload, they naively process the  queries in the query workload individually, independent of each other. There is a need for index structures that can improve the performance of a query workload by taking into consideration the important characteristics of the workload. In this paper, we present our proposed index structure, \textit{qwLSH}, that can efficiently process similarity search query workloads in high-dimensional spaces. The following are the contributions of this paper:
\begin{itemize}[leftmargin=*]
	\item Given the important observation, that different data have different query processing needs, we intelligently divide a given cache for efficient processing of a query workload
	\item We present novel cost models for intelligently dividing a cache given a query workload
	\item We present different cache utilization strategies for LSH-based query workloads
	\item Finally, we use real datasets to show the efficiency of our proposed index structure, \textit{qwLSH}, by comparing against state-of-the-art algorithms.
\end{itemize} 

To the best of our knowledge, there are no existing works that particularly tackle the problem of solving query workloads in high-dimensional spaces in an efficient manner. In this paper, we present the design and analysis of \textit{qwLSH} that was particularly designed to efficiently execute query workloads in high-dimensional spaces. The paper is organized as follows: in Section \ref{sec:relWork}, we present the related works. In Section \ref{sec:prelim}, we present the key concepts and preliminaries necessary to understand the problem domain and \textit{qwLSH}. In Sections \ref{sec:probSpec} and \ref{sec:qwLSH}, we formalize the problem statement and explain the design of \textit{qwLSH} respectively. We present our experimental analysis in Section \ref{sec:eval} and finally conclude in Section \ref{sec:concl}. 

\section{Related Work}
\label{sec:relWork}
Similarity search queries in high-dimensional spaces have wide-ranging applications in various domains such as multimedia retrieval, artificial intelligence, etc. Exact tree-based index structures (such as R-trees, KD-trees, etc.) work efficiently for low-dimensional data, but as the dimensions increase ($\sim>10$), their performance degrades and are often outperformed by brute-force linear scans (a well-known problem called the \textit{curse of dimensionality}) \cite{Indyk:1998:ANN:276698.276876}. 
%In order to alleviate the effects of \textit{curse of dimensionality}, 
Approximate techniques were proposed where the performance of the queries was drastically improved by trading off some accuracy of the query result. 
%This trade-off is often decided by a user-input success probability ($1-\delta$); the higher the target success probability, the higher the accuracy of the query result at the expense of the query performance, and vice-versa. 
Locality Sensitive Hashing, first proposed in \cite{Indyk:1998:ANN:276698.276876}, is a popular approximate technique for solving similarity search queries in high-dimensional spaces. 
%The key idea behind LSH is to map high-dimensional data to lower-dimensional spaces, that are generated using random projections, such that data points that are nearby in the original high-dimensional space map to the same buckets in the lower-dimensional space with a high probability. Conversely, points that are far apart in the original space should only be mapped to the same bucket with a small probability. 
The original work was proposed to solve the (R, c)-Near Neighbor problem, which is a decision-based version of the Approximate Nearest Neighbor problem \cite{Datar:2004:LHS:997817.997857}. While the LSH family (explained further in Section \ref{sec:prelim}) in \cite{Indyk:1998:ANN:276698.276876} was originally defined for the Hamming distance, it was later defined for other distance measures such as the Euclidean distance \cite{Datar:2004:LHS:997817.997857}. 

%\subsection{Effective Variants of LSH}
\noindent \textbf{Effective Variants of LSH: }
LSH has been shown to be useful in various domains such as biomedical sciences \cite{Buhler2001, Berlin2015, Rasheed2013}, geological sciences \cite{Yoon2015, Buaba2010}, etc. Several works were subsequently proposed \cite{Lv:2007:MLE:1325851.1325958, Bawa:2005:LFS:1060745.1060840, Liu:2014:SEI:2732939.2732947, Gan:2012:LHS:2213836.2213898, Sun:2014:SSC:2735461.2735462, Huang:2015:QLH:2850469.2850470, Tao:2010:EAN:1806907.1806912, Gao:2015:SHC:2783258.2783284} to improve upon the original work. 
%The original LSH work suffered from three main drawbacks: (1) it required large number of hash functions and hash layers in order to return the required number of results, (2) the number of candidates returned were large and hence the required IO cost was high, and (3) the performance of the LSH index was dependent on several parameters, and if the parameters were not chosen correctly, the accuracy and/or the performance would be drastically affected. In order to alleviate these problems, several works were subsequently proposed \cite{Lv:2007:MLE:1325851.1325958, Bawa:2005:LFS:1060745.1060840, Liu:2014:SEI:2732939.2732947, Gan:2012:LHS:2213836.2213898, Sun:2014:SSC:2735461.2735462, Huang:2015:QLH:2850469.2850470, Tao:2010:EAN:1806907.1806912, Gao:2015:SHC:2783258.2783284}. 
In \cite{Bawa:2005:LFS:1060745.1060840}, the authors created a prefix-tree of hash functions for each hash layer to appropriately decide on the number of hash functions to use during query processing in order to return the desired number of top-$k$ results. In \cite{Lv:2007:MLE:1325851.1325958}, the authors propose a probing technique to look into neighboring buckets of the query point's hash bucket, and hence instead of creating more number of layers to get the desired number of top-$k$ results, their technique probes into neighboring buckets. In \cite{Tao:2010:EAN:1806907.1806912}, the authors represent each of the points in the lower dimensional space using Z-order (also called Morton) codes which are then further retrieved effectively by finding the closeby points based on the distances between their Z-order codes. While these techniques improve upon the original LSH and its most popular implementation, E2LSH\footnote{\url{https://www.mit.edu/~andoni/LSH/}}, they still suffer from having large index sizes and slow processing speeds. Recently, in \cite{Gan:2012:LHS:2213836.2213898}, the authors presented C2LSH, in which they proposed two novel ideas: a ``collision counting" approach that counted the number of times a candidate object is mapped to the same bucket and a ``virtual rehashing" approach that automatically incremented the lookup space in each projection without having the need to physically rehash the data. 
%C2LSH was shown to be very effective in reducing the index size and the query processing speed. 
%while removing the need to input important parameters, such as the radius of the query and width of the hash bucket (which are data dependent parameters), and satisfying the target success probability. 
In \cite{Gao:2015:SHC:2783258.2783284}, the authors analyzed the data distribution and created indexes of different granularity to make their index structure scalable for different data with different distributions. 
%Intuitively, this approach improves the ratio slightly but also introduces an overhead of maintaining and accessing the B-trees. 
In \cite{Sun:2014:SSC:2735461.2735462}, a projection-based method that uses only 6 to 10 random projections is presented.  While the index size is much smaller than \cite{Gan:2012:LHS:2213836.2213898} and \cite{Huang:2015:QLH:2850469.2850470}, the accuracy is much worse and it is shown to be unstable \cite{Huang:2015:QLH:2850469.2850470}. In \cite{Liu:2014:SEI:2732939.2732947}, the authors present SK-LSH, where they propose a novel linear order on the index files stored on the external storage based on their Z-order codes. The idea is that if nearby index files are stored contiguously on the external storage, then the number of I/Os will be reduced. 
%In our experiments, we found that their work (and subsequently their code) not only requires the user to input the radius of the hash bucket and the number of necessary hash layers (which are data dependent), but also consistently produces higher number of candidates than C2LSH and QALSH. But most interestingly, they do not have the input target success probability ($1-\delta$) in their formulation, making it difficult to perform a fair comparison with other LSH-based works (which all have the target success probability as an input). 
In \cite{Huang:2015:QLH:2850469.2850470}, the authors build upon the ``collision counting'' and the ``virtual rehashing'' approaches presented in \cite{Gan:2012:LHS:2213836.2213898} by proposing to create ``query-aware'' hash functions. QALSH builds B-trees on every hash function, and then given a query object, perform range searches on each of the B-trees. This method has shown to be the most accurate and fast LSH-based technique \cite{Gao:2015:SHC:2783258.2783284}. 

%\subsection{Query Workloads}
\noindent \textbf{Query Workloads: }
Many index structures and systems have been particularly designed to execute query workloads efficiently \cite{Curino:2010:SWA:1920841.1920853, Aly:2015:AAQ:2831360.2831361, Tzoumas:2009:WIC:1687627.1687761, DBLP:journals/pvldb/NagarkarCB15, DBLP:conf/edbt/NagarkarC14, Quamar:2013:SSW:2452376.2452427}. \cite{Curino:2010:SWA:1920841.1920853, Pavlo:2012:SAD:2213836.2213844} uses data partitioning techniques that are query workload-aware on 1D data. In \cite{Achakeev:2012:SQL:2396761.2398577, Tzoumas:2009:WIC:1687627.1687761, Aly:2015:AAQ:2831360.2831361, Aly:2016:KWP:2835776.2835841}, the authors present different query workload-aware partitioning techniques for existing data structures, such as R-trees, on 2D spatial data. In \cite{DBLP:conf/edbt/NagarkarC14} and \cite{DBLP:journals/pvldb/NagarkarCB15}, the authors present unique cost models to bring in the most relevant part of the index structures into the main memory for future use based on past query workload statistics for 1D and 2D space respectively. The above works deal with data partitioning techniques for 1D and 2D data. To the best of our knowledge, there is no existing work that deals with improving the efficiency of query workloads in high-dimensional spaces. 

%\subsection{Cache-Conscious Querying}
\noindent \textbf{Cache-Conscious Querying: }
There have been several works that design techniques to reduce cache misses to improve the query performance \cite{10.1007/978-3-319-56111-0_1, Rao:2000:MBT:342009.335449, Hankins:2003:ENS:781027.781063}. An experimental survey on cache conscious algorithms can be found at \cite{Yotov:2007:ECC:1248377.1248394}. In the domain of LSH and its variants, there is one main work that has attempted making LSH cache-conscious \cite{Sundaram:2013:SSS:2556549.2556574}. The authors present a caching strategy to improve the index construction cost. Their main goal is to speed up indexing (and searching) over streaming Twitter data \textit{in a distributed environment}. In the caching strategy presented in their work, the authors present a hierarchical 2-level hashing approach to reduce cache misses during the index construction phase. During query processing, the authors utilize the cache by ``software prefetching", i.e., by prefetching succeeding data items into the cache. Our cache utilization strategies are very different than the techniques used in \cite{Sundaram:2013:SSS:2556549.2556574}. Our strategies are based on the important observation as explained in Section \ref{sec:motiv}, and our main goal is to speed up the processing speed of the entire query workload, which is different from \cite{Sundaram:2013:SSS:2556549.2556574}. 

\section{Key Concepts and Preliminaries}
\label{sec:prelim}

In this section, we present the key concepts underlying LSH. We rely mainly on the terminologies presented in E2LSH\footnote{\url{https://www.mit.edu/~andoni/LSH/}}, C2LSH \cite{Gan:2012:LHS:2213836.2213898}, and QALSH \cite{Gao:2015:SHC:2783258.2783284}. 

Let $\mathcal{D}$ be a database of $n$ data objects in a $d$-dimensional Euclidean space $\mathcal{R}^d$. Let $\lVert o_1, o_2\rVert$ denote the Euclidean distance between two objects $o_1$ and $o_2$. Given a query object $q$ in $\mathcal{R}^d$, the $c$-ANN search (for an approximation ratio $c>1$) returns all objects $o \in \mathcal{D}$ such that $\lVert o, q\rVert \leq c\times \lVert o^*, q\rVert$, where $o^*$ is the true nearest neighbor of $q$ in $\mathcal{D}$. Similarly, the $c$-$k$-ANN problem finds the top-$k$ objects $o_i \in \mathcal{D}$ where $1 \leq i \leq k$, and such that $\lVert o_i, q\rVert \leq c\times \lVert o_i^*, q\rVert$, where $o_i^*$ is the true $i$th nearest neighbor of $q$. 

\begin{definition}[A Locality Sensitive Hashing Family]
	A hash function family $\mathcal{H}$ is said to be $(r,	c, P1, P2)$-sensitive if it satisfies all following conditions for
	any two points $x$ and $y$ in a dataset $\mathcal{D} \subset \mathcal{R}^d$:
\end{definition}

\begin{itemize}
	%	\vspace*{-0.02in}	
	\item if $|x - y| \leq r$, then $Pr[h(x) = h(y)] \geq P_1$, and
	%	\vspace*{-0.02in}	
	\item if $|x - y| > cr$, then $Pr[h(x) = h(y)] \leq P_2$
\end{itemize}

Here, $c$ is an approximation ratio, $P_1$ and $P_2$ are probabilities, $r$ is the distance between two points commonly referred to as the radius, and in order for the definition to work, $c > 1$ and
$P_1 > P_2$. The above definition states that the two points $x$ and $y$ are hashed to the same bucket with a very high
probability $\geq P_1$ if they are close to each other (i.e.  the distance between the two points is less than or equal to $r$), and if they are not close to each other (i.e. the distance between the two points is greater than $cr$), then they will be hashed to the same bucket with a low probability $\leq P_2$. 

In the original LSH scheme for Euclidean distance, each hash function is defined as $h_{a,b} (v) = {\frac{a.v + b}{w}},$
where $a$ is a $d$-dimensional random vector with entries chosen independently from the standard normal distribution and $b$ is a real
number chosen uniformly from $[0, w)$, such that $w$ is the width of the hash bucket \cite{Datar:2004:LHS:997817.997857}. 
This leads to the following collision probability function~\cite{Liu:2014:SEI:2732939.2732947}:\\
\begin{equation}
\label{eqn:origp1}
P(r) = \int_{0}^{w}{\frac{1}{r} \frac{2}{\sqrt{2\pi}} e^{\frac{-t^2}{2r^2}}(1-\frac{t}{w})dt}.
\end{equation}
Note that the collision probability is governed by the width, $w$, of
the hash bucket: if the size is chosen to be much larger than the query
radius, then there can be a lot of candidates generated.  If the size
is chosen to be much smaller than the query radius, then there can be
potentially several misses.

%\subsection{Key Differences between Basic LSH and State-of-the-art techniques}
%\label{sec:lshdiff}

\subsection{Collision Counting}

C2LSH \cite{Gan:2012:LHS:2213836.2213898} introduced the technique of ``collision counting''
because of which it is not necessary to have $l$ hash layers. Instead of hash layers, the index requires $m$ hash functions and a \textit{collision count threshold} to find the candidate points. From here on, since each hash function can be viewed as a single hash layer, we use the terms hash layers and hash functions interchangeably. 

Recent LSH variants (such as C2LSH \cite{Gan:2012:LHS:2213836.2213898}, QALSH \cite{Huang:2015:QLH:2850469.2850470}, etc.) have an upper bound on the number of candidate objects returned, i.e., the number of data points that have to be brought from the external storage into the main memory. This bound is often controlled by a user-input, ``allowed" false positives, $v$. Intuitively, higher the $v$, less time is needed to find the candidate objects, and vice-versa. As a good trade-off, these works set $v = 100$, which is what we continue using in our work as well. Thus, for a maximum $k$ (i.e. the desired number of output results) of 100, the maximum number of data objects that will be read from the disk will be 200 (since $k + v = 200$). The basic LSH formulation does not have an upper bound on the number of candidate points that are needed to be accessed from the external storage.

\section{Problem Specification}
\label{sec:probSpec}

In this paper, our goal is to create an efficient index structure to execute similarity search query workloads in high-dimensional spaces. 
Let us consider we are given a query workload $\mathcal{Q}$ that consists of $q$ queries, where $\forall q \subset \mathcal{D}$. For each query $q \in \mathcal{Q}$, given $k$ and an approximation ratio $c>1$, we return the top-$k$ results such that $\lVert o_i, q\rVert \leq c\times \lVert o_i^*, q\rVert$, where $o_i^*$ is the true $i$th nearest neighbor of $q$. 

As mentioned in Section \ref{sec:motiv}, our goal is to leverage the important observation that datasets with different cardinalities and dimensionalities have different IO costs, and then design the cache such that it is most effectively utilized for a given query workload. 

Given a cache $C$, our goal is to divide the cache into two parts: $C_I$ and $C_D$, where $C_I$ is the part of the cache that stores the index files (i.e. data from the hash buckets) and $C_D$ is the part of the cache that stores the data objects, such that $size(C) = size(C_I) + size(C_D)$. Assume $cost(C)$ denotes the cost of bringing all necessary files (index files + data objects) into the cache $C$. Similarly, $cost(C_I)$ denotes the cost of bringing the index files into the cache, and $cost(C_D)$ denotes the cost of bringing the data objects into the cache, such that $cost(C) = cost(C_I) + cost(C_D)$. In order to understand the design of \textit{qwLSH}, we need to first understand the costs, $cost(C_I)$ and $cost(C_D)$, in detail. 

\noindent\boldmath{$cost(C_I)$: }\unboldmath Let us first consider the cost of bringing all necessary index files into $C_I$ for a single query, $q$. Note that, we have $m$ projections in our index. Let us assume that the cost for bringing necessary index files for $i$th projection is: $cost(C_{I_i}^q)$. Also, the amount of data that is read from the index files is a function of the cardinality of the dataset (Figure \ref{fig:motivationCard}). Let us denote this constant as $\alpha_{card}$. Thus, we have: 
\begin{equation}
cost(C_I) = \alpha_{card} \times \sum_{i=1}^{m}(cost(C_{I_i}^q))
\end{equation}

\noindent\boldmath{$cost(C_D)$: }\unboldmath Similarly, let us first consider the cost of bringing all necessary data objects into $C_D$ for a single query, $q$. Note that, as explained in Section \ref{sec:prelim}, for a single query, the worst case scenario would be to have to read $k + v$ data objects from the external storage. Let us denote this as $w$ (i.e. $w = k + v$). Let us denote the cost of reading a single $j$th data object as $cost(C_{D_j}^q)$. The size of each data object is a function of the dimensionality of the dataset (Figure \ref{fig:motivationDim}). Let us denote this constant as $\alpha_{dim}$. Thus, we have:
\begin{equation}
cost(C_D) = \alpha_{dim} \times \sum_{j=1}^{w}(cost(C_{D_j}^q))
\end{equation}

\noindent Note that the above costs are for a single query. When we have more than 1 query in the given query workload $\mathcal{Q}$, assuming that there will be at least 1 index file that will be reused within the queries, $cost(C_I)$ will depend on the size of $C_I$. If a large amount of $C$ is dedicated to $C_I$, then the $cost(C_I)$ will be less and vice-versa. Conversely, $cost(C_D)$ will depend on the size of $C_D$. 
Thus, given a query workload $\mathcal{Q}$, we want to find $size(C_I)$ (or $size(C_D)$) such that $cost(C)$ is minimized:
%
%\begin{equation*}
%\begin{aligned}
%& {\text{minimize}}
%& & \mathlarger{\sum_{g=1}^{|\mathcal{Q}|}}\bigg(\alpha_{card} \times \sum_{i=1}^{m}(cost(C_{I_i}^g)) \\
%& & + \alpha_{dim} \times \sum_{j=1}^{w}(cost(C_{D_j}^g))\bigg) \\
%& \text{subject to}
%& & size(C_I) = size(C) - size(C_D).
%\end{aligned}
%\end{equation*}

\begin{eqnarray*}
	%	\begin{split}
	{\text{minimize}} & & \mathlarger{\sum_{g=1}^{|\mathcal{Q}|}}\bigg(\alpha_{card} \times \sum_{i=1}^{m}(cost(C_{I_i}^g)) \\
	& + &  \alpha_{dim} \times \sum_{j=1}^{w}(cost(C_{D_j}^g))\bigg) \\
	\text{subject to}
	& & size(C_I) = size(C) - size(C_D).
	%	\end{split}
\end{eqnarray*}%}%

% HOW TO DO SUB-FIGURES HORIZONTALLY
%\begin{figure*}
%	\centering
%	\begin{subfigure}[b]{0.48\textwidth}
%		\centering
%		{\setlength{\fboxsep}{0pt}
%			\setlength{\fboxrule}{0.2pt}
%			\fbox{\includegraphics[width=\linewidth]{Figures/IndexIOSize}}
%		}
%		\caption{Total Size of Index Files Read}
%	\end{subfigure}\quad
%	\begin{subfigure}[b]{0.48\textwidth}
%		\centering
%		{\setlength{\fboxsep}{0pt}
%			\setlength{\fboxrule}{0.2pt}
%			\fbox{\includegraphics[width=\linewidth]{Figures/DataIOSize}}
%		}
%		\caption{Total Size of Data Files Read}
%	\end{subfigure}
%	\caption{The total size of Index and Data Files read for different datasets with different cardinalities and dimensionalities. The datasets are synthetically generated with uniform distribution. The bars with striped patterns show predicted values since we could not generate indexes for those datasets due to memory limitations.}\label{fig:indDataModel}
%\end{figure*}

\begin{figure}
	\centering
	\begin{subfigure}[b]{0.18\textwidth}
		\centering
		\includegraphics[width=\linewidth]{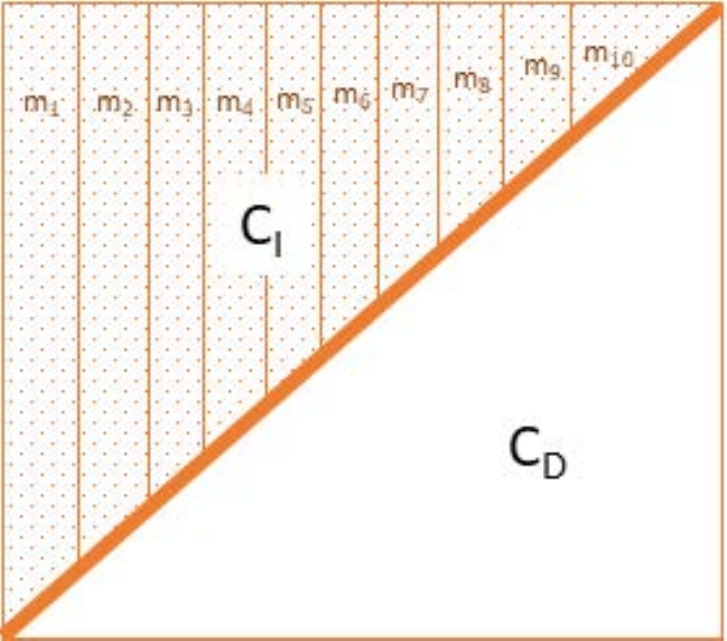}
		\caption{Strategy 1}
		\label{fig:strategy1}
	\end{subfigure}\hfill
	%		\begin{subfigure}[b]{0.22\textwidth}
	%			\centering
	%			\includegraphics[width=\linewidth]{Figures/PSLSHDesign3-2.pdf}
	%			\caption{When $\rho$ = 2}
	%			\label{fig:rho2}
	%		\end{subfigure}\hfill
	\begin{subfigure}[b]{0.18\textwidth}
		\centering
		\includegraphics[width=\linewidth]{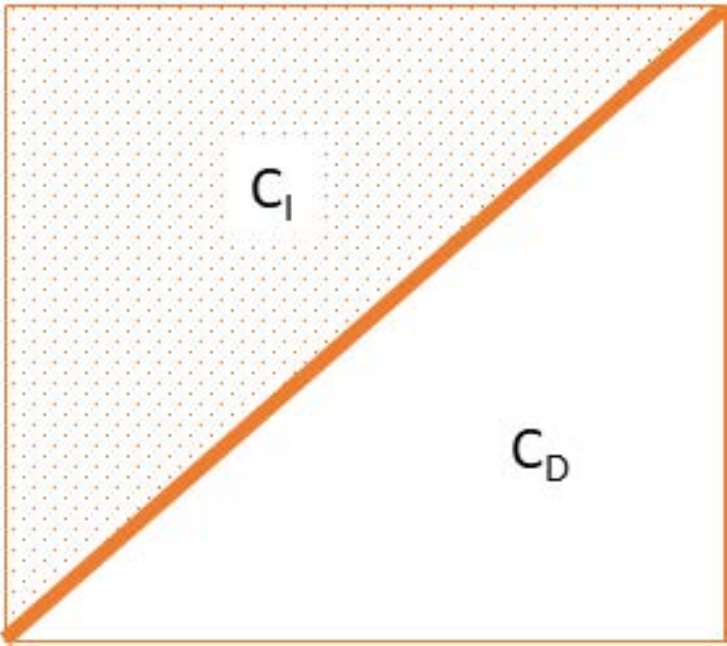}
		\caption{Strategy 2}
		\label{fig:strategy2}
	\end{subfigure}
	\caption{Different Strategies for designing the cache}
	\label{fig:qwlsh}
\end{figure}

\begin{figure}
	\centering
	\includegraphics[width=0.95\linewidth]{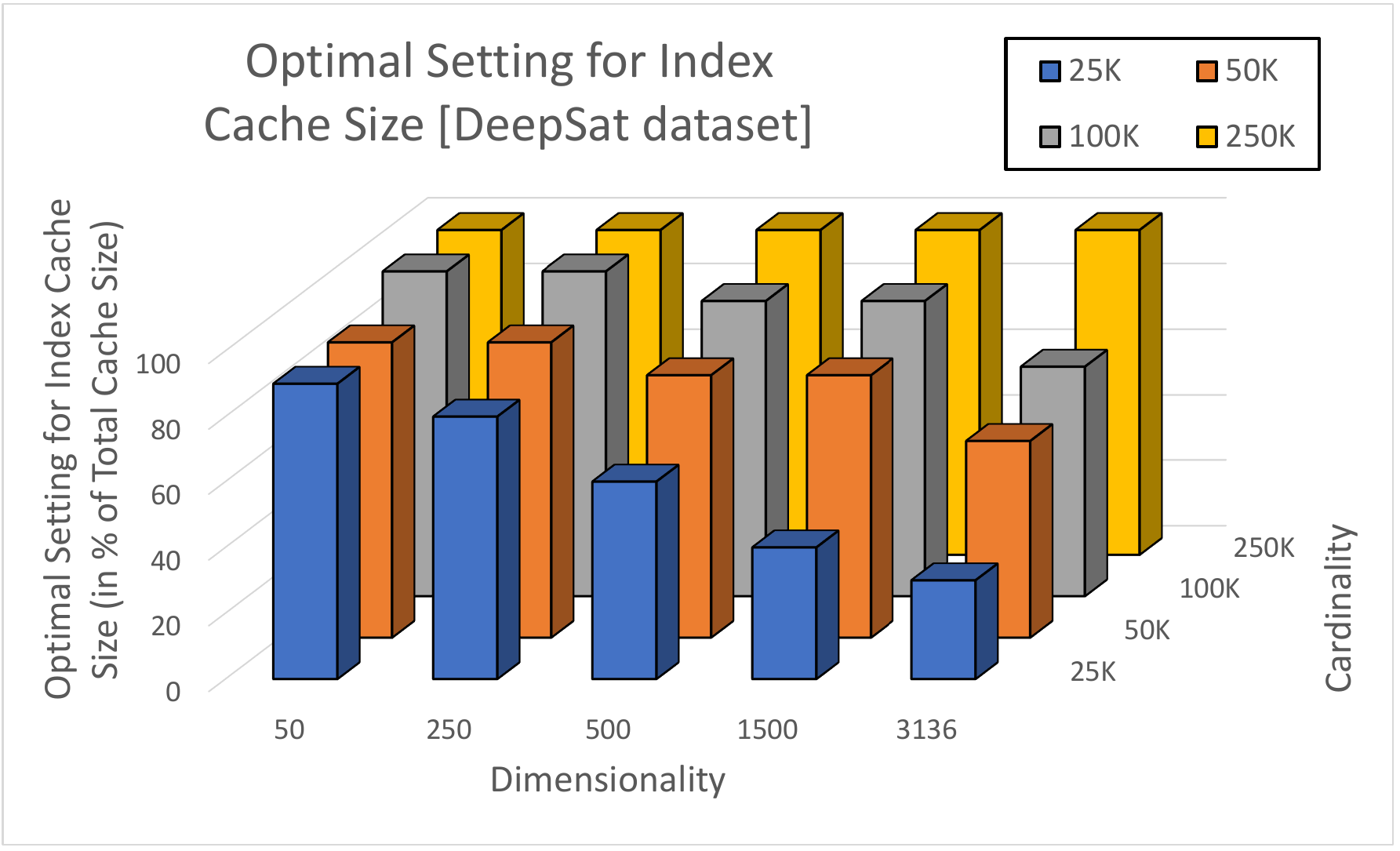}
	\caption{Optimal Setting for Index Cache Size (in \% of the Total Available Cache Size) on different versions of the Deepsat dataset [16MB Cache Size, |Q|=250]}
	\label{fig:costModelDefPar}
\end{figure}

\begin{figure}
	\centering
	\includegraphics[width=0.8\linewidth]{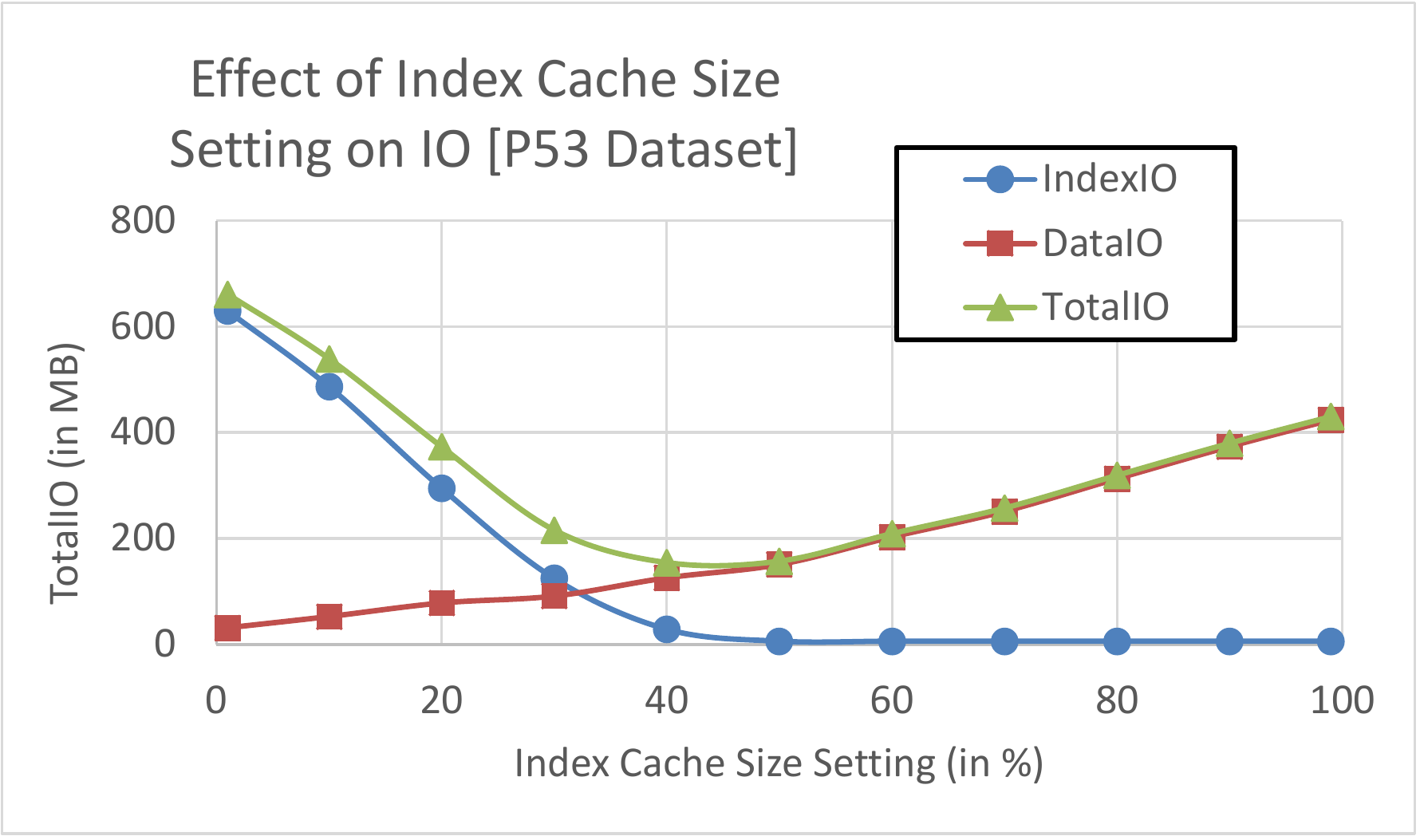}
	\caption{Effect of Index Cache Size Setting on the IndexIO, DataIO, and TotalIO for P53 dataset [16MB Cache Size, |Q|=250]}
	\label{fig:P53_IndexIO_DataIO_Comparison}
\end{figure}

\section{qwLSH}
\label{sec:qwLSH}

In this section, we describe our proposed index structure, \textit{qwLSH}. Given a query workload, our goal is to intelligently divide the cache based on the cardinality and dimensionality of the given dataset. In order to do this, we leverage the detailed cost model presented in Section \ref{sec:probSpec}. In this section, we will present the two strategies for dividing the cache, along with unique cost models.

\noindent\textbf{Naive Solution:} In LSH (and its variants), the cache is not intelligently divided into two parts: $C_I$ and $C_D$. For a given query workload, the cache (using the default cache replacement policy, such as MRU) will treat the index files and the data object files equal, regardless of the cardinality and the dimensionality of the dataset. As shown in Figures \ref{fig:motivationCard} and \ref{fig:motivationDim}, the cardinality and the dimensionality of the dataset affects the total size of index files that are read into the cache and the total size of data objects respectively. We leverage this observation that naive solutions do not. In Section \ref{sec:eval}, we show the benefit of leveraging this observation into designing the cache intelligently. 

\subsection{Design of qwLSH}
\label{sec:design}

For each query, LSH index structures access each of the $m$ hash functions in order to find the candidate data object IDs. Once the candidate data object IDs are found for the query, the data objects are brought from the external storage in order to remove the false positives. Given the total cache size, our goal is to find how much of the total cache size to allocate to the index cache, $C_I$, and the data cache, $C_D$. Once we determine the appropriate split between $C_I$ and $C_D$, we use the popular cache replacement policy, MRU, to decide which index files need to be evicted from $C_I$ and which data files need to be evicted from $C_D$. With the intuition that queries in query workloads are often times near each other \cite{Aly:2015:AAQ:2831360.2831361, DBLP:journals/pvldb/NagarkarCB15}, we choose to use MRU. Before we present the cost models that we used to determine the appropriate split between $C_I$ and $C_D$, 
we first present two strategies based on how the index files are accessed and cached.

\noindent\textbf{Strategy 1:} In this strategy, the index cache, $C_I$, is further divided uniformly between the $m$ hash functions, as shown in Figure \ref{fig:strategy1}. In each of the \textit{sub-index cache}, we have an MRU replacement policy specific to the \textit{sub-index cache}. 

\noindent\textbf{Strategy 2:} In this strategy, the index cache, $C_I$, is not further divided uniformly between the $m$ hash functions, as shown in Figure \ref{fig:strategy2}. We have a single MRU replacement policy for the index cache that stores (and if needed, evicts) index files from different hash functions. 

\noindent In both strategies, there is a separate MRU replacement policy for the data cache, $C_D$. The intuition behind Strategy 1 is that each projection will use the cache uniformly. On the other hand, the intuition behind Strategy 2 is that some projections might require bringing more hash buckets into the cache than others (which would be the case when the data distribution is skewed or the usage of individual projections is different). 

\subsection{Cost Models of qwLSH}
We have seen in Figures \ref{fig:motivationCard} and \ref{fig:motivationDim} that the cardinality and dimensionality of the dataset affects the size of index files and data files that need to be read into the cache respectively. In order to determine the appropriate split between the index cache, $C_I$, and the data cache, $C_D$, we train a model based on the size of index files and data files that are read from the external storage for different settings (cardinality and dimensionality) on the DeepSat dataset. Figure \ref{fig:costModelDefPar} shows the optimal setting of the index cache size for different dataset characteristics that returns the least amount of total IO. This model validates our observation that different data characteristics utilize the cache differently. For instance, datasets with low dimensionality (=50) require more index cache utilization ($\geq$80\%) since the data points are small in size. On the contrary, datasets with low cardinality ($\leq$100K) and high dimensionality ($\geq$500) require less index cache utilization since the data points are large in size and the number of index files needed to be brought in memory are low. Hence, it is beneficial to cache more data points than index files to reduce the total amount of IO. 

Figure \ref{fig:P53_IndexIO_DataIO_Comparison} shows the effect of different Index Cache size setting on the IndexIO, DataIO, and the TotalIO for P53 dataset. As the Index Cache size setting increases (i.e. more \% of the cache is allocated to the index cache), as expected, the DataIO cost increases while the IndexIO cost decreases. When Index Cache size setting is at 40\%, the system incurs the least amount of IO while processing the entire query workload. The model presented in Figure \ref{fig:costModelDefPar} also estimates the Index Cache size setting for P53 dataset to be at 40\%. Our model is dependent on the 
cache size and the number of queries in the query workload. We show in Section \ref{sec:eval} that our novel index structure can still adapt for different cache sizes and number of queries. 

\subsection{Workflow of qwLSH}
In order to find the optimal index cache size for efficient utilization of the cache, we have to first generate a training model that shows the behavior of different cache sizes for different dataset settings (Figure \ref{fig:costModelDefPar}). Note that, this process is done offline and does not require the knowledge of the query workloads beforehand. During query processing, once we know the characteristics of the dataset (i.e. the cardinality and dimensionality), we refer to our model to determine the index cache size and the data cache size. Note that, the underlying LSH index (that is stored on the disk) is not changed. 
In \textit{qwLSH}, we use a linked list (to decide which objects to evict or store in the cache) and an unordered hash map (for fast retrieval of the objects). In Section \ref{sec:eval}, we show that this overhead is minimal and the gains achieved from our novel models outweigh the cost of the overhead.

\begin{table}[t]
	\centerline{
		\begin{tabular}{|c|c|}\hline
			{\bf Parameter} & {\bf Value range} \\ \hline\hline 
			\# of Queries in Query Workload ($|\mathcal{Q}|$) & 50;\; 100;\; {\bf 250};\;\\ \hline 
			Total Cache Size (in MB) & 8;\;{\bf 16};\;20;\; \\ \hline
		\end{tabular}
	}
	\caption{Parameters and Default Values (in bold)}\label{tab:paramPSLSH}
\end{table}

\begin{figure}
	\centering
	\includegraphics[width=0.8\linewidth]{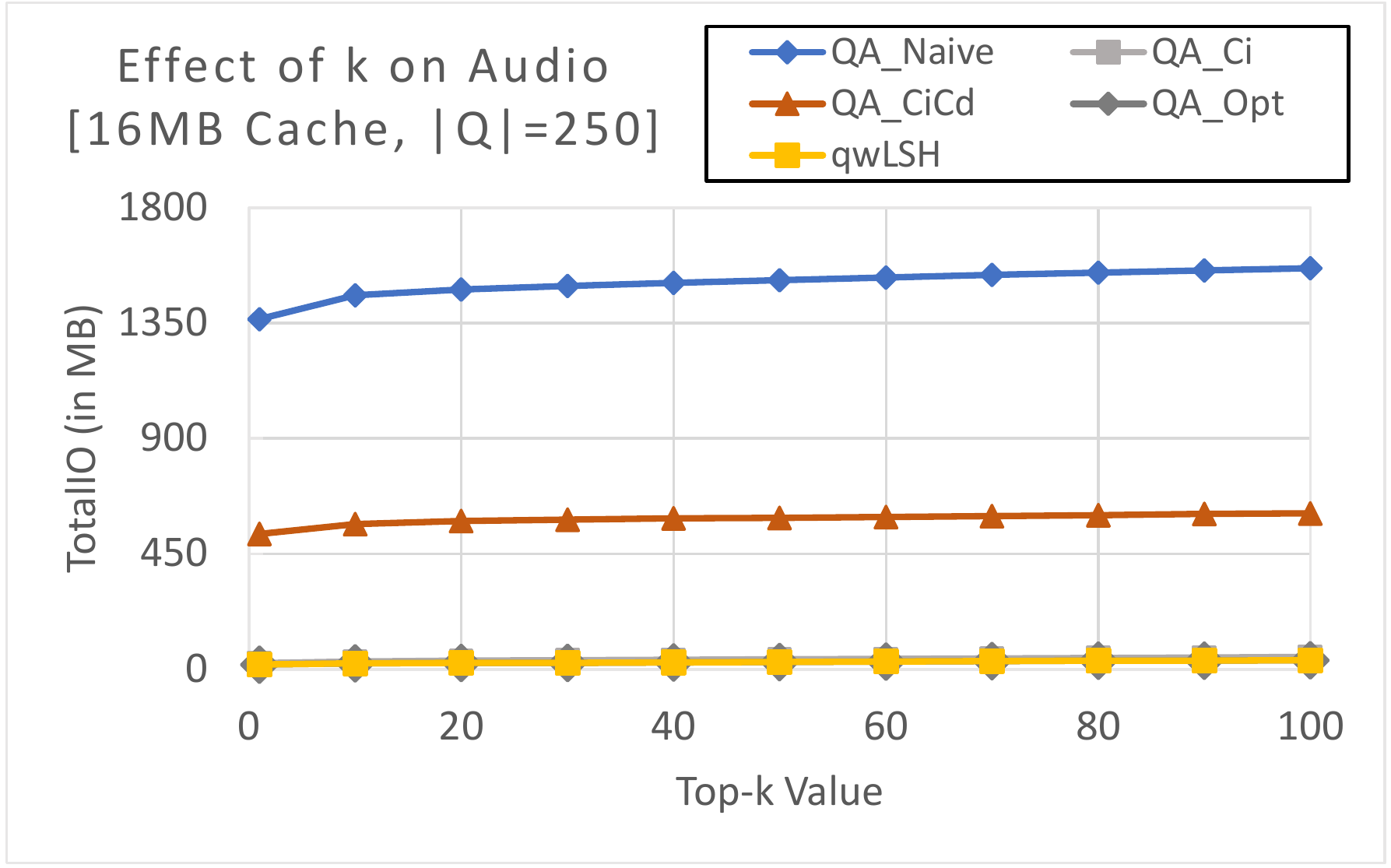}
	\caption{Effect of varying top-k on Audio dataset [16MB Cache, |Q|=250]}
	\label{fig:kAudio}
\end{figure}

\begin{figure}
	\centering
	\includegraphics[width=0.8\linewidth]{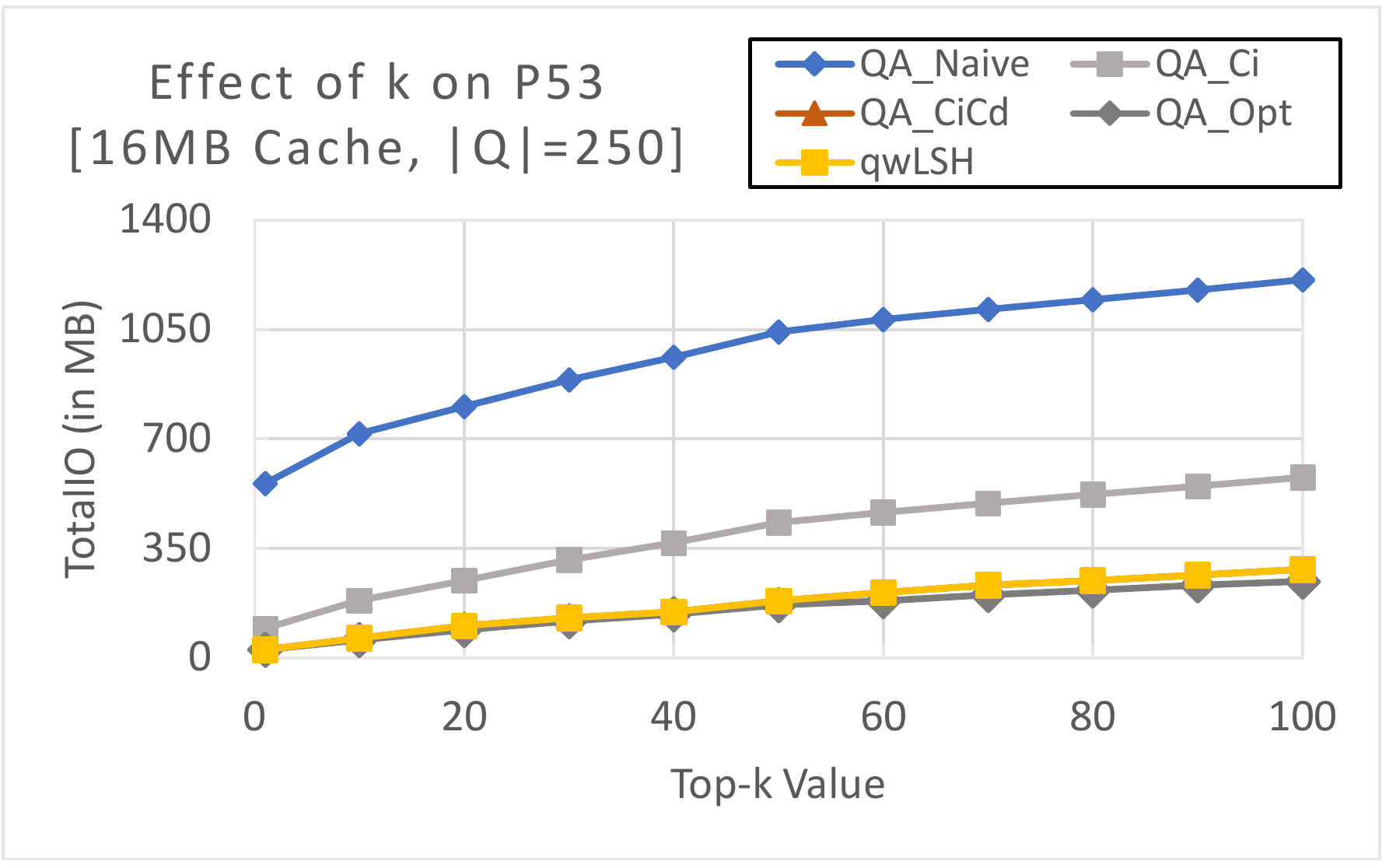}
	\caption{Effect of varying top-k on P53 dataset [16MB Cache, |Q|=250]}
	\label{fig:kP53}
\end{figure}

\begin{figure*}
	\centering
	\begin{subfigure}[b]{0.23\textwidth}
		\centering
		{\setlength{\fboxsep}{0pt}
			\setlength{\fboxrule}{0.2pt}
			\fbox{\includegraphics[width=\linewidth]{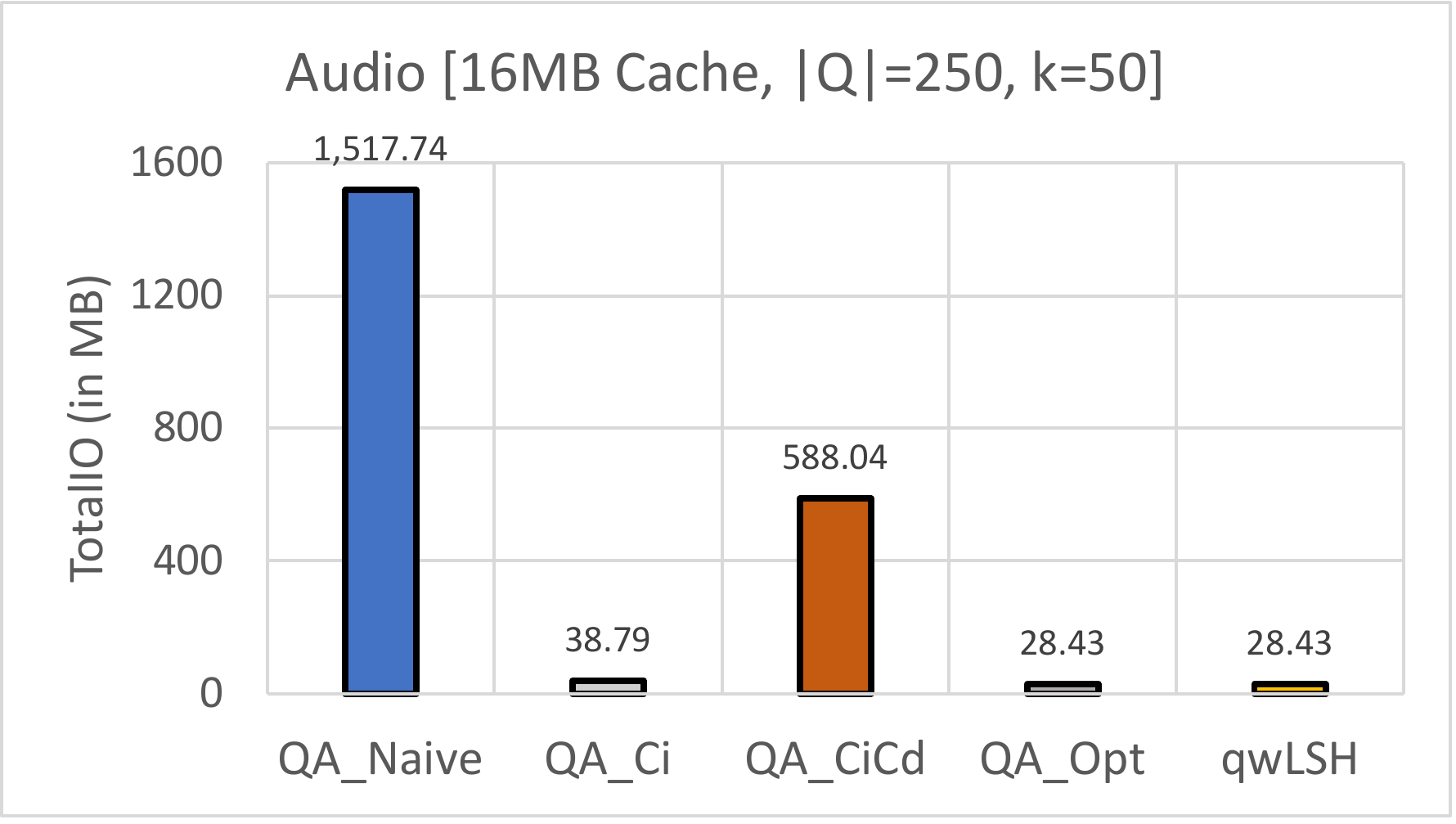}}
		}
	\end{subfigure}\quad
	\begin{subfigure}[b]{0.23\textwidth}
		\centering
		{\setlength{\fboxsep}{0pt}
			\setlength{\fboxrule}{0.2pt}
			\fbox{\includegraphics[width=\linewidth]{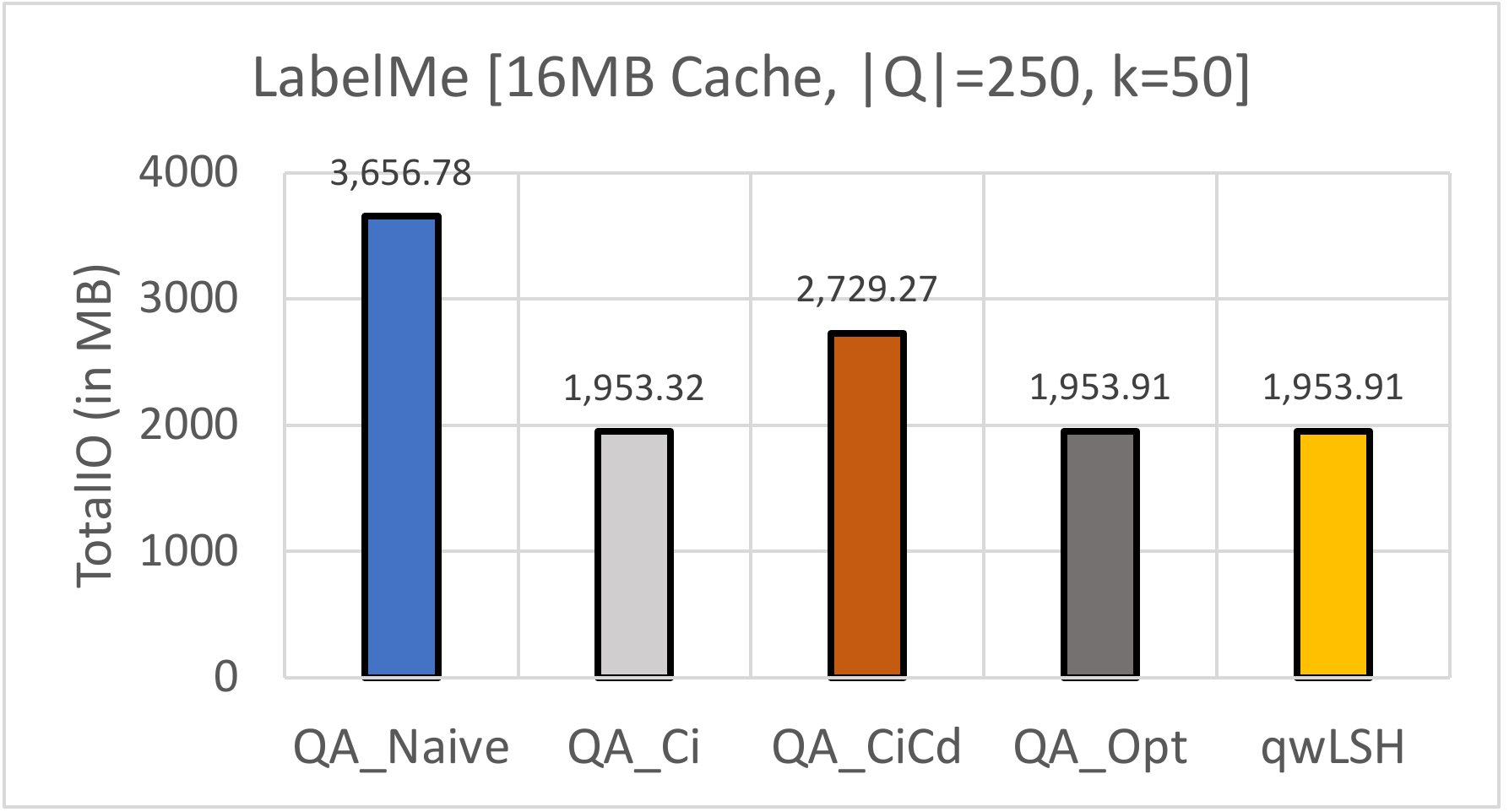}}
		}
	\end{subfigure}\quad
	\begin{subfigure}[b]{0.23\textwidth}
		\centering
		{\setlength{\fboxsep}{0pt}
			\setlength{\fboxrule}{0.2pt}
			\fbox{\includegraphics[width=\linewidth]{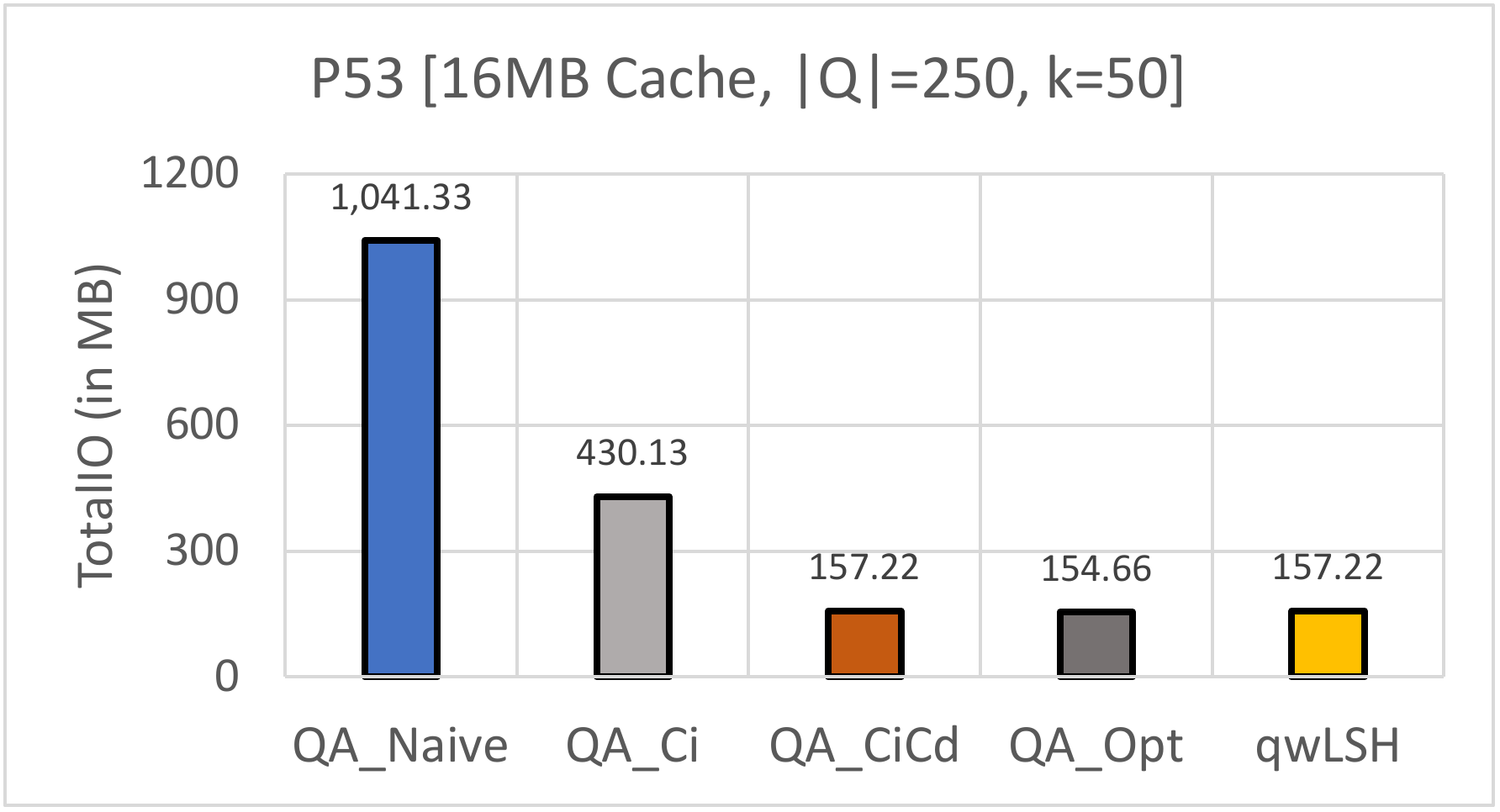}}
		}
	\end{subfigure}\quad
	\begin{subfigure}[b]{0.23\textwidth}
		\centering
		{\setlength{\fboxsep}{0pt}
			\setlength{\fboxrule}{0.2pt}
			\fbox{\includegraphics[width=\linewidth]{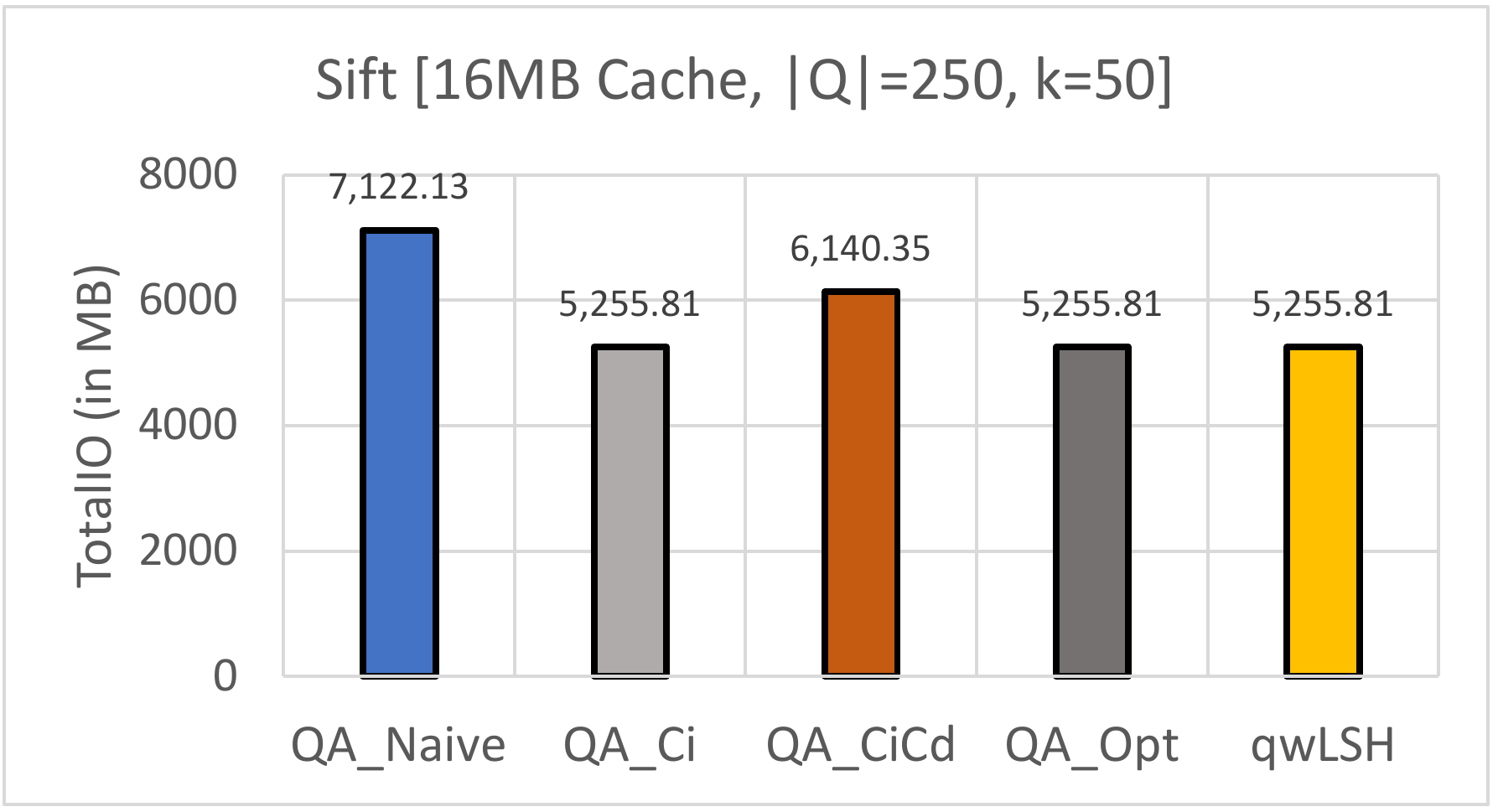}}
		}
	\end{subfigure}
\vspace*{-0.15in}
	\caption{Comparison of qwLSH ({\bf TotalIO}) against its alternatives (for different real datasets) and default settings}\label{fig:defaultSettings}
%	\vspace*{-0.15in}
\end{figure*}

\begin{figure*}
	\centering
	\begin{subfigure}[b]{0.23\textwidth}
		\centering
		{\setlength{\fboxsep}{0pt}
			\setlength{\fboxrule}{0.2pt}
			\fbox{\includegraphics[width=\linewidth]{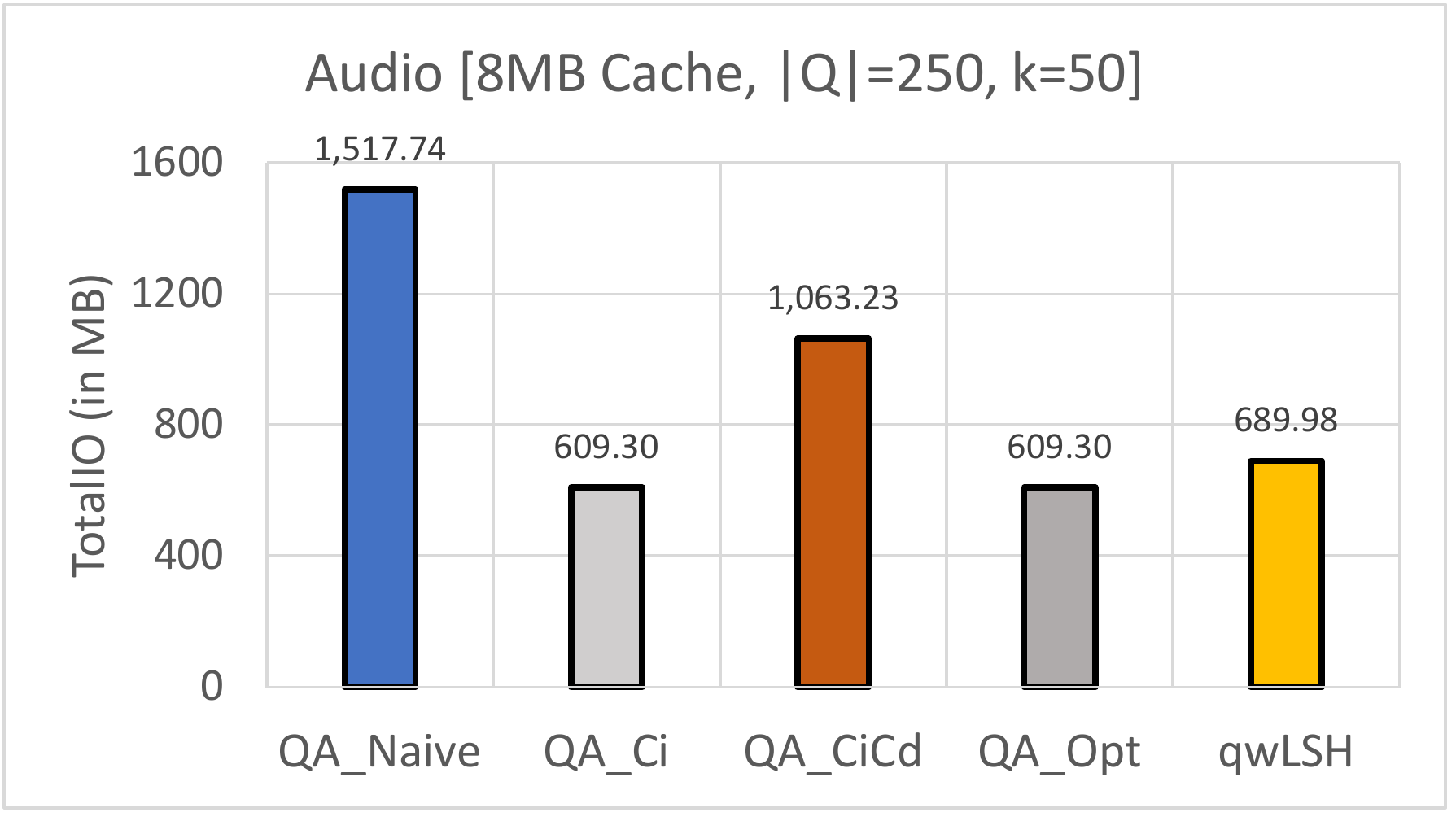}}
		}
	\end{subfigure}\quad
	\begin{subfigure}[b]{0.23\textwidth}
		\centering
		{\setlength{\fboxsep}{0pt}
			\setlength{\fboxrule}{0.2pt}
			\fbox{\includegraphics[width=\linewidth]{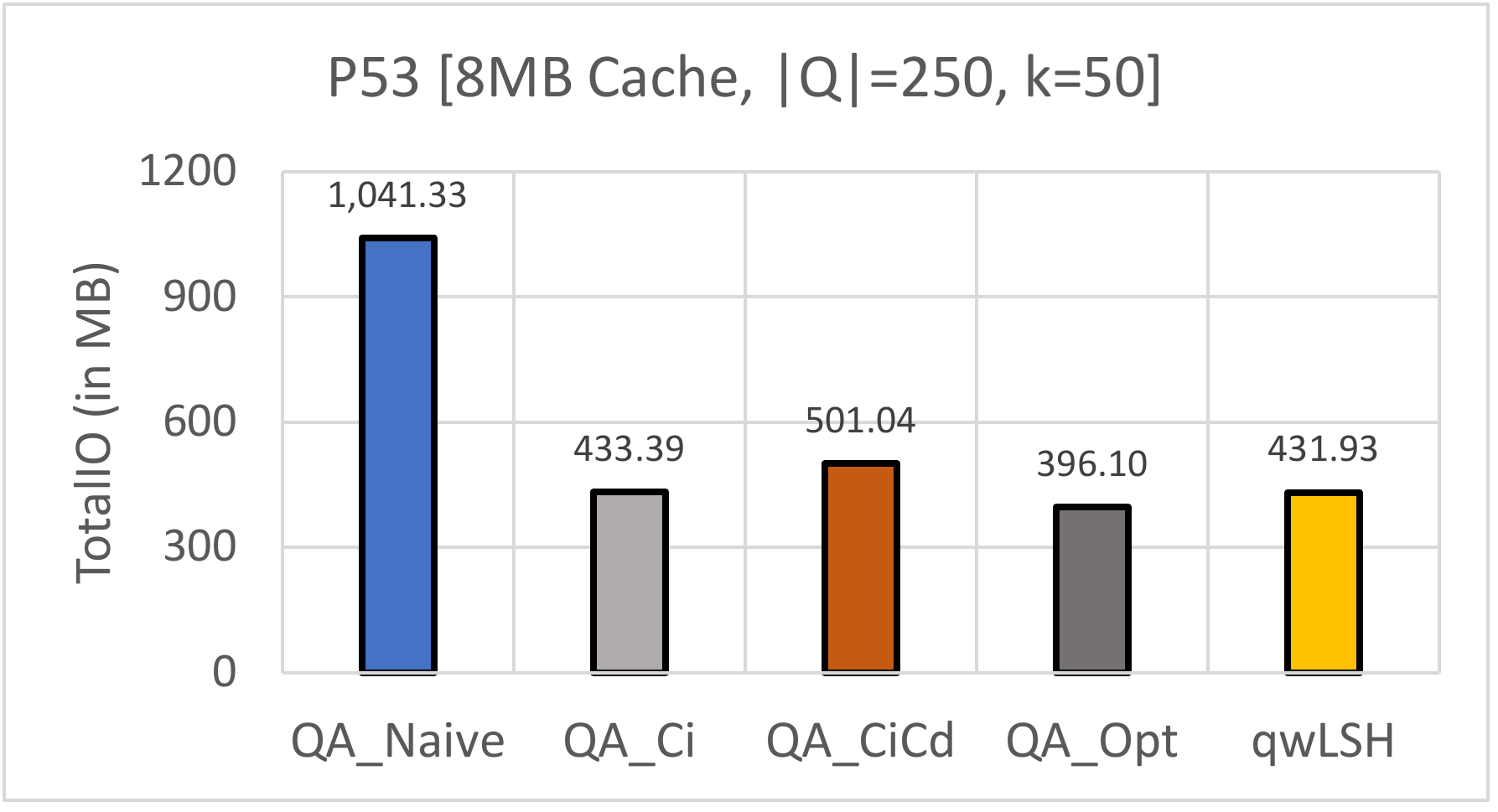}}
		}
	\end{subfigure}\quad
	\begin{subfigure}[b]{0.23\textwidth}
		\centering
		{\setlength{\fboxsep}{0pt}
			\setlength{\fboxrule}{0.2pt}
			\fbox{\includegraphics[width=\linewidth]{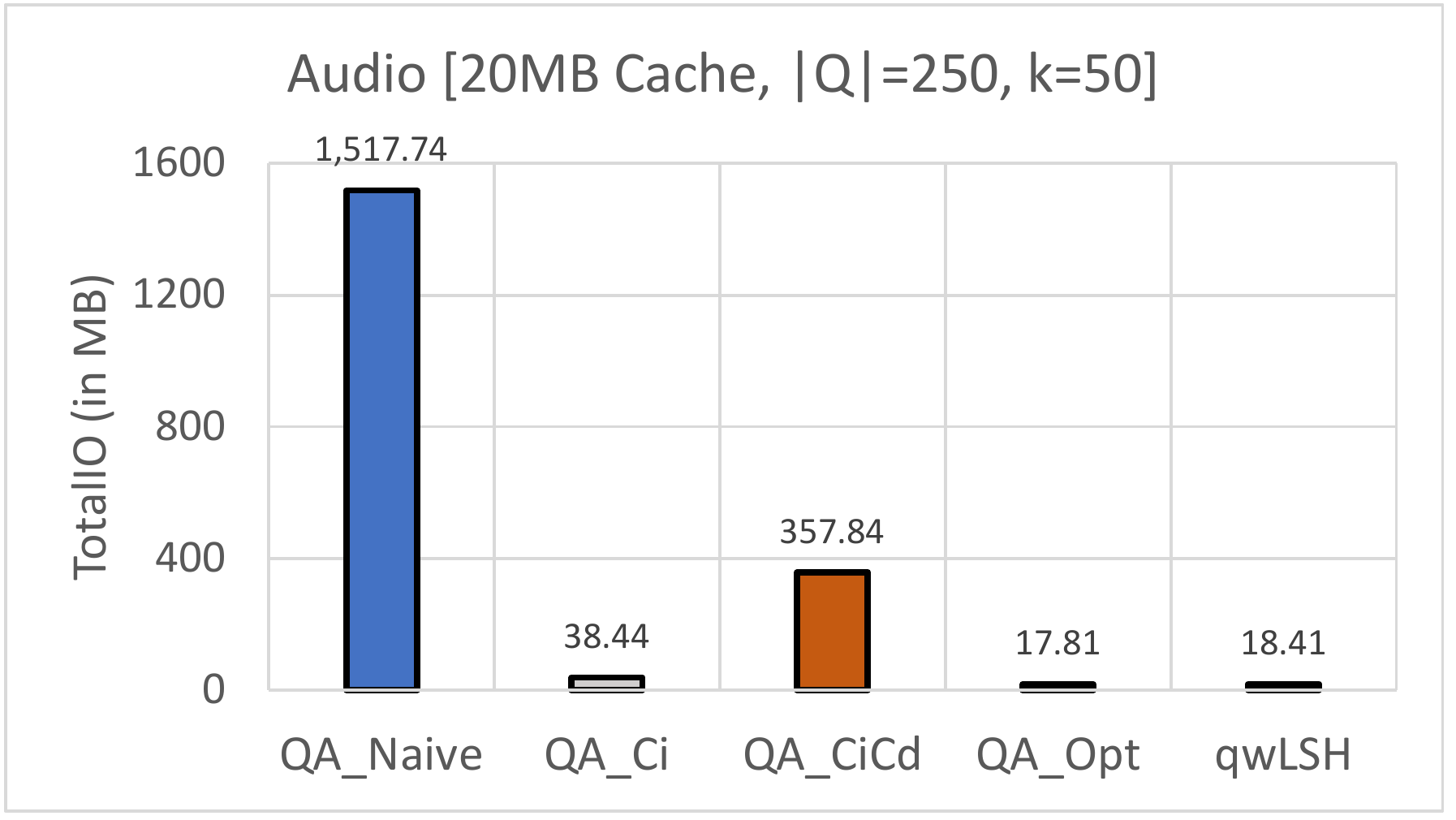}}
		}
	\end{subfigure}\quad
	\begin{subfigure}[b]{0.23\textwidth}
		\centering
		{\setlength{\fboxsep}{0pt}
			\setlength{\fboxrule}{0.2pt}
			\fbox{\includegraphics[width=\linewidth]{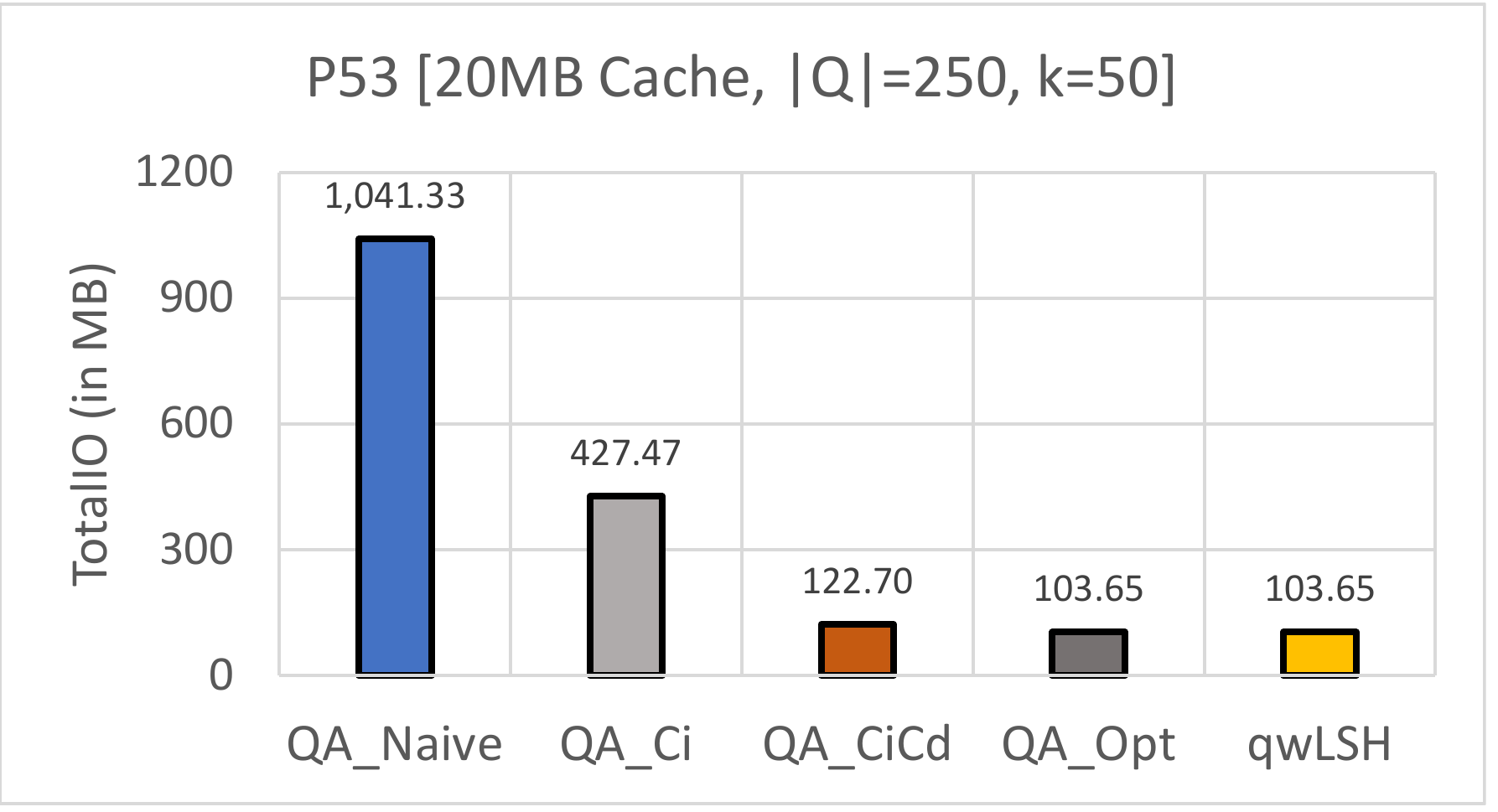}}
		}
	\end{subfigure}
\vspace*{-0.15in}
	\caption{Comparison of qwLSH ({\bf TotalIO}) for varying cache size against its alternatives (for Audio and P53 datasets)}\label{fig:diffCacheSize}
\end{figure*}

\begin{figure*}
	\centering
	\begin{subfigure}[b]{0.23\textwidth}
		\centering
		{\setlength{\fboxsep}{0pt}
			\setlength{\fboxrule}{0.2pt}
			\fbox{\includegraphics[width=\linewidth]{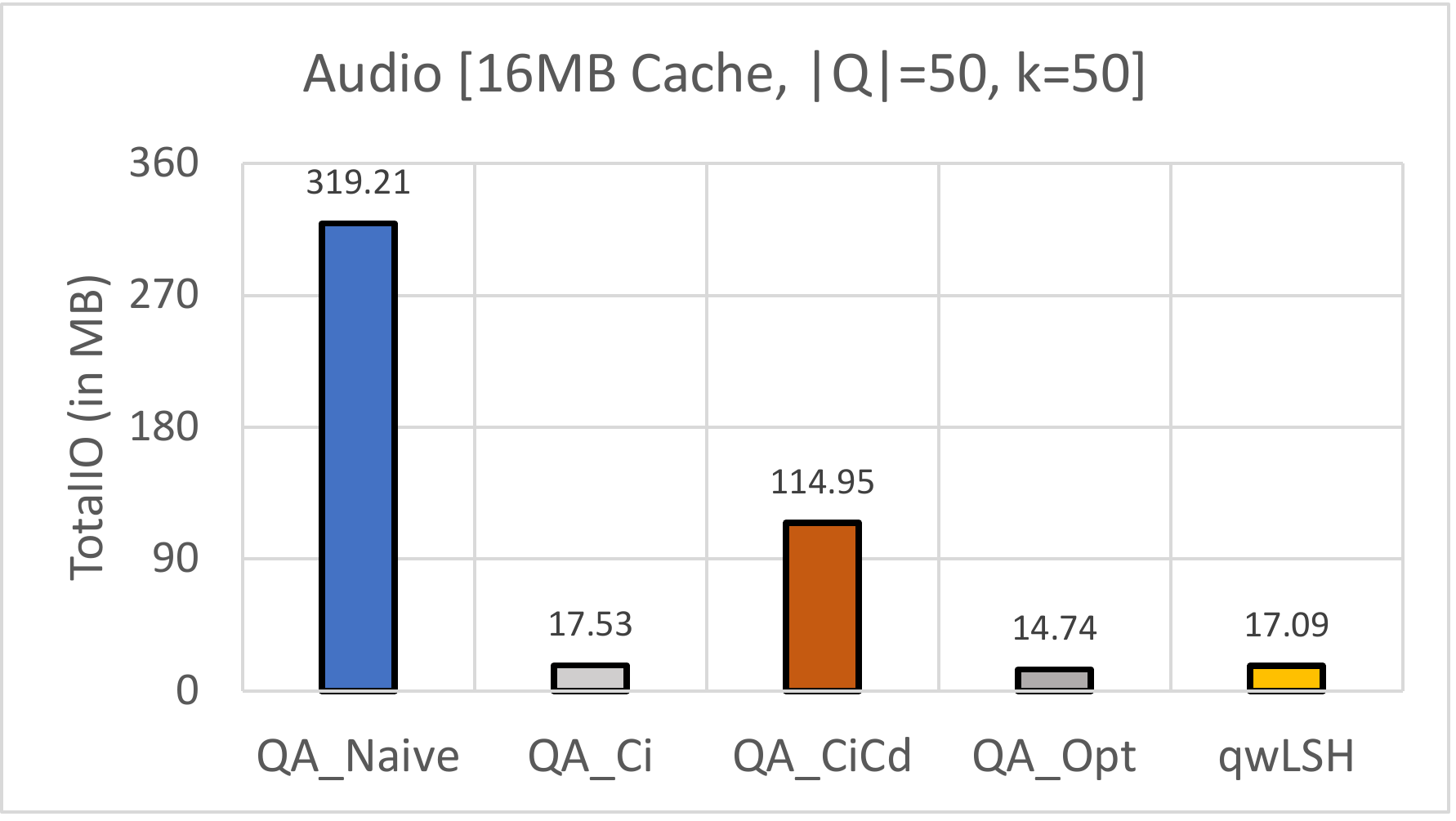}}
		}
	\end{subfigure}\quad
	\begin{subfigure}[b]{0.23\textwidth}
		\centering
		{\setlength{\fboxsep}{0pt}
			\setlength{\fboxrule}{0.2pt}
			\fbox{\includegraphics[width=\linewidth]{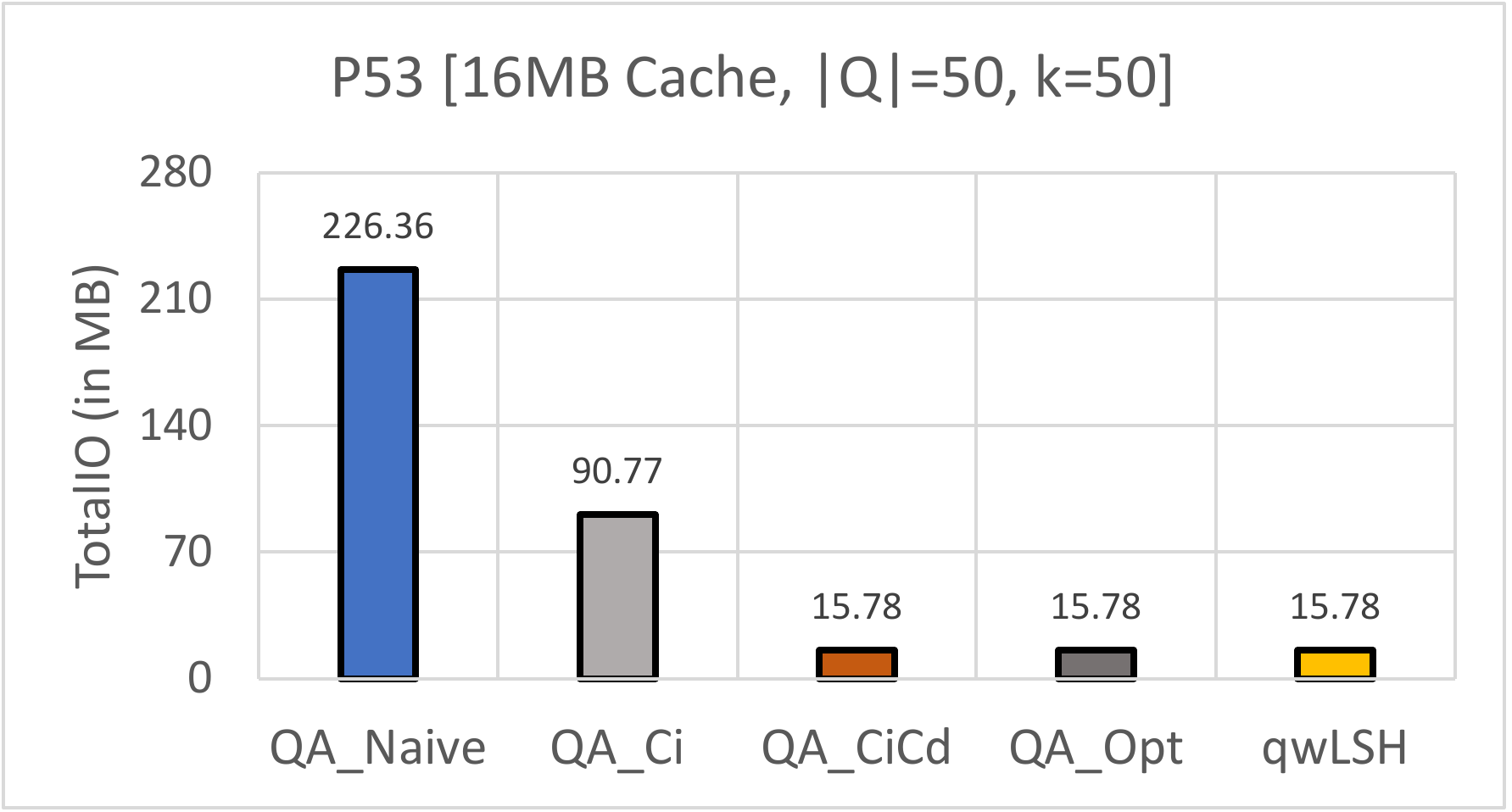}}
		}
	\end{subfigure}\quad
	\begin{subfigure}[b]{0.23\textwidth}
		\centering
		{\setlength{\fboxsep}{0pt}
			\setlength{\fboxrule}{0.2pt}
			\fbox{\includegraphics[width=\linewidth]{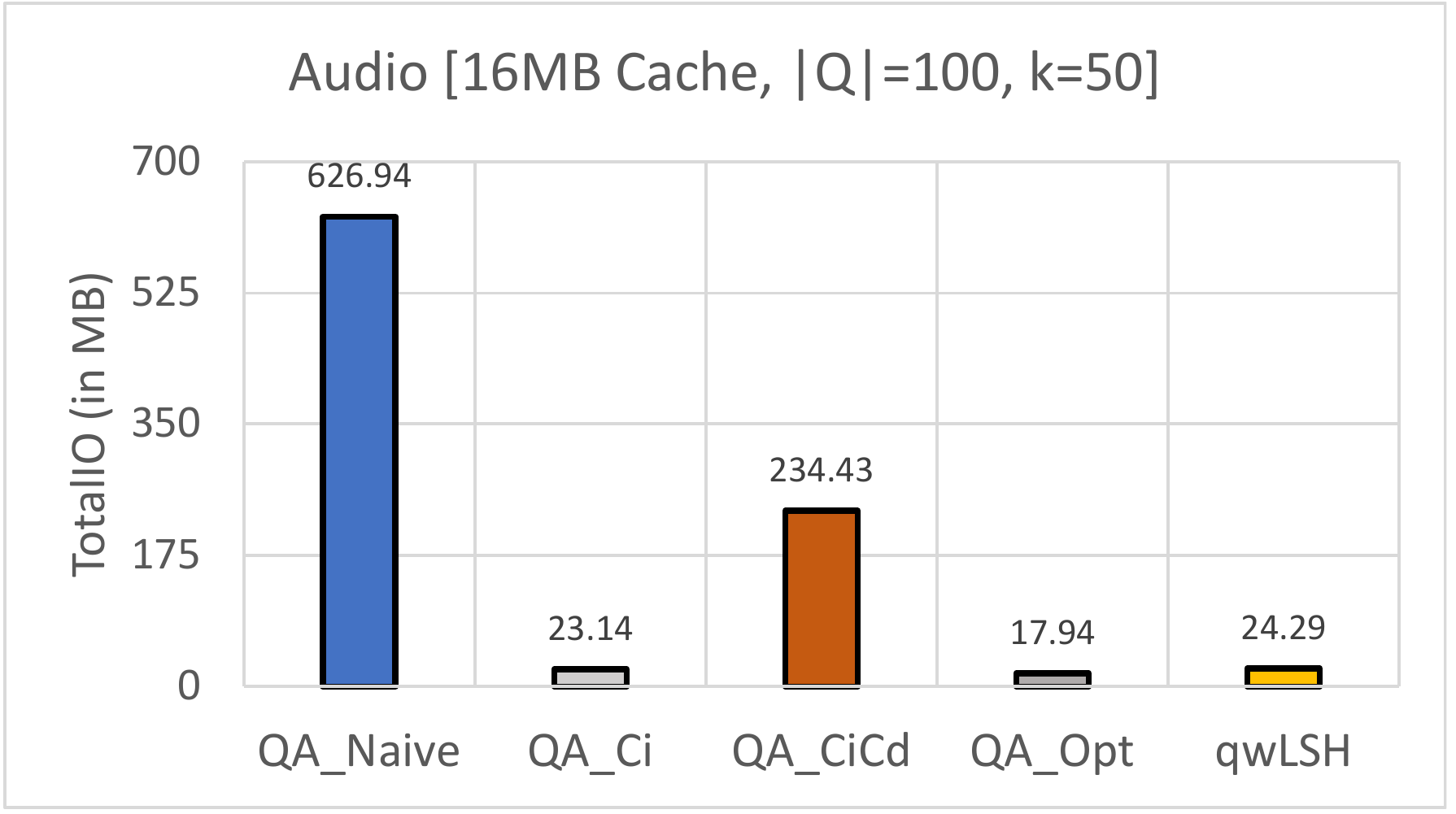}}
		}
	\end{subfigure}\quad
	\begin{subfigure}[b]{0.23\textwidth}
		\centering
		{\setlength{\fboxsep}{0pt}
			\setlength{\fboxrule}{0.2pt}
			\fbox{\includegraphics[width=\linewidth]{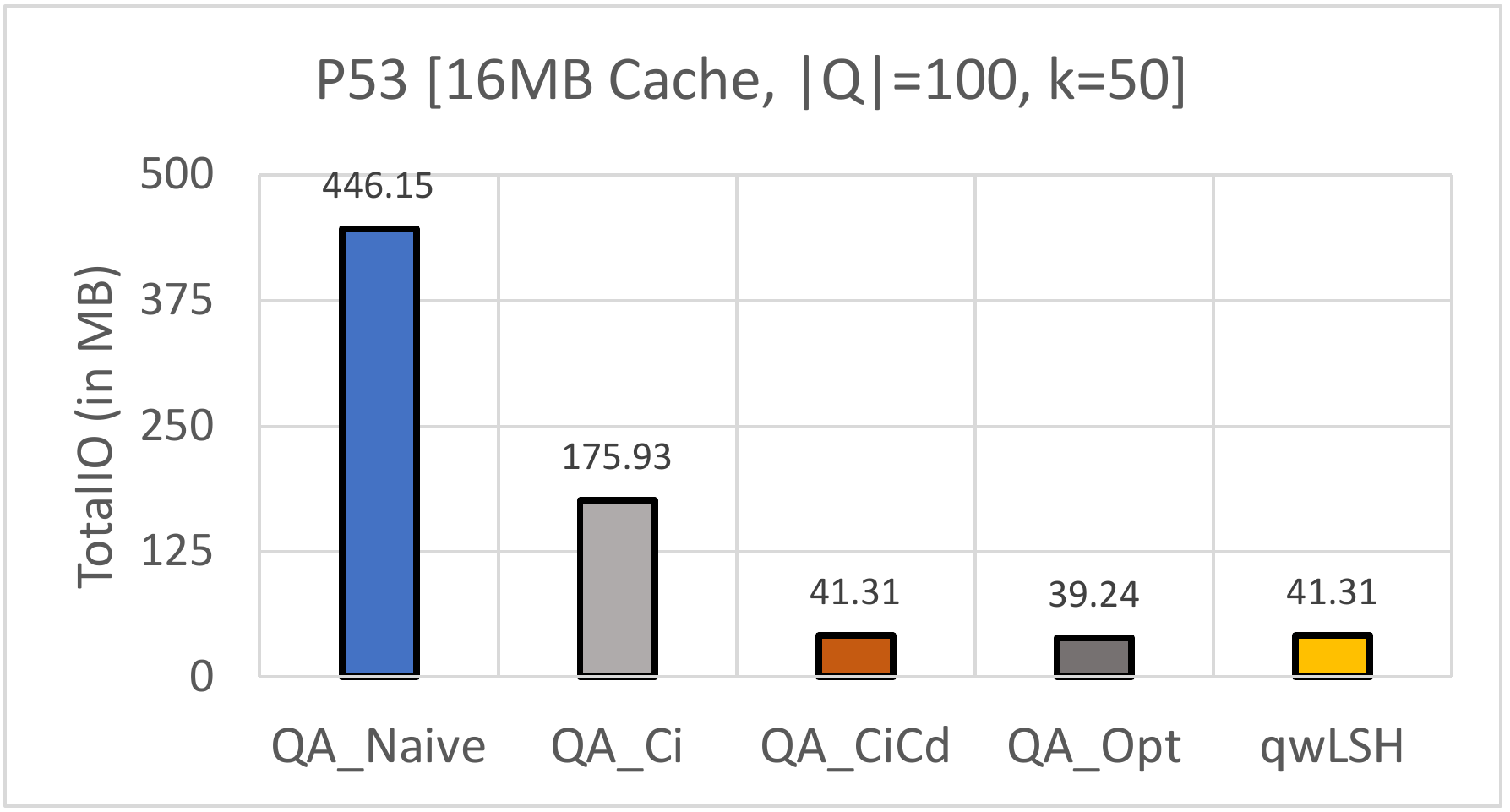}}
		}
	\end{subfigure}
\vspace*{-0.15in}
	\caption{Comparison of qwLSH ({\bf TotalIO}) for varying number of queries against its alternatives (for Audio and P53 datasets)}\label{fig:diffQueries}
\end{figure*}

\section{Experimental Evaluation}
\label{sec:eval}
In this section, we evaluate the effectiveness of our proposed index structure, \textit{qwLSH}. For these evaluations, we use several real data sets with different cardinality and dimensionality, under different system parameters. All experiments were run on machines with the following specifications: Intel Core i7-6700, 16GB RAM, 2TB HDD, and Ubuntu 16.04 operating system. The reported results are an average of 5 runs. We used the state-of-the-art QALSH \cite{Gao:2015:SHC:2783258.2783284} as our base implementation\footnote{\textit{qwLSH} can be implemented over any state-of-the-art LSH variant.}. All codes were written in C++-11.
% and compiled with gcc v5.4 with the -O0 optimization flag. 
We implement a cache in the existing QALSH code. For all following alternatives, we use the default settings of QALSH ($c=2$, $w=2.719$, $\delta=1/e$) that are mentioned in \cite{Gao:2015:SHC:2783258.2783284}. 

\noindent Since there is no work that directly aims at solving our problem, we compare our work with the following alternatives: 
\begin{itemize}[leftmargin=*]
	\item \textbf{QALSH\_Naive:} We compare against the baseline QALSH algorithm. 
	\item \textbf{QALSH\_Ci:} In this alternative, we allocate the maximum size of 99\% (of the total cache size) to $C_I$. Hence, $C_D$ will have 1\% of the total cache size allocation.	
	\item \textbf{QALSH\_Cd:} Here, we allocate the maximum size of 99\% (of the total cache size) to $C_D$. Hence, $C_I$ will have 1\% of the total cache size allocation. 		
	\item \textbf{QALSH\_CiCd:} In this alternative, we equally divide the cache into $C_I$ and $C_D$, i.e. we allocate 50\% (of the total cache size) to $C_I$ and 50\% to $C_D$. 
	\item \textbf{QALSH\_Opt:} We try 11 different settings for $C_I$ ($C_I=1, 10, 20,...,90, 99$) and report the most efficient setting. 	
	
\end{itemize}

\vspace*{-0.15in}
\subsection{Datasets}
In order to train our models, we needed a large dataset with high-cardinality and high-dimensionality. Due to the lack of such datasets, we followed the same technique used to generate the Mnist\footnote{\url{http://yann.lecun.com/exdb/mnist/}} dataset by downloading 324,000 images from the Deepsat dataset (that consists of airborne images of different land surfaces)\footnote{\url{https://www.kaggle.com/crawford/deepsat-sat6/home}}. Each image has 28*28 pixels. For each pixel, we store the RGB and Near-Infrared value, and hence for each image, we get a 28*28*4=3136-dimensional point. From this dataset, we generated 20 different datasets (with different cardinalities and dimensionalities) to create our model (Figure \ref{fig:costModelDefPar}). 
In order to test our model and the effectiveness of \textit{qwLSH}, for our experiments, we used the following four commonly used real high-dimensional datasets: 
\begin{itemize}[leftmargin=*]
	
	\item \textbf{Audio\footnote{\url{http://www.cs.princeton.edu/cass/audio.tar.gz}}}  This dataset consists of 54387 192-dimensional points. It consists of human-labled sound clips.	
	\item \textbf{LabelMe\footnote{\url{http://labelme.csail.mit.edu/Release3.0/browserTools/php/dataset.php}}}  This dataset consists of 181093 512-dim. points. 
%	\item \textbf{Mnist\footnote{\url{http://yann.lecun.com/exdb/mnist/}}}  This dataset consists of 60000 50-dimensional points. It is used to represent handwritten digits.
	\item \textbf{P53\footnote{\url{https://archive.ics.uci.edu/ml/datasets/p53+Mutants}}}  This dataset consists of 31008 5409-dimensional points. Since our test dataset consisted of only 3136 dimensions, we reduce the dimensionality of each point to 3000.
	\item \textbf{Sift\footnote{\url{http://corpus-texmex.irisa.fr/}}}  This dataset consists of 250,000 128-dim. points. 
	
\end{itemize}

\subsection{Evaluation Criteria and Parameters}

Note that, our goal is not to improve the accuracy of the queries in the query workload, but to improve the total runtime of the query workload. Since we do not change the logic of the LSH algorithm, the accuracy of each query executed in \textit{qwLSH} is the same as the accuracy of the underlying LSH algorithm. Hence, due to space limitations, we do not report the accuracy of the individual queries. For that, we ask the reader to refer to the QALSH paper \cite{Gao:2015:SHC:2783258.2783284}. 
We evaluate the effectiveness of \textit{qwLSH} by comparing the size of the data (i.e. the index files and the data objects) that is needed to be brought into the cache. We do not report the index size or the index construction cost, since they would be the same as the underlying LSH implementation that we use, which in our case is QALSH \cite{Gao:2015:SHC:2783258.2783284}. We focus on one specific criterion: size of the data that is needed to be brought into the cache. 

\noindent On our test datasets, we found that Strategy 1 and Strategy 2 (Section \ref{sec:design}) gave similar results (mostly because of the fact that QALSH treats each projection as the same, i.e., the number of hash buckets brought into the main memory is the same for each projection). Due to space constraints, we only evaluate Strategy 1 and compare with the alternatives presented earlier in this section. 

\noindent Table \ref{tab:paramPSLSH} shows our default parameter settings and the different ranges that we consider. We generate query workloads by choosing queries from dense regions of the dataset (that reflects our motivation and prior works \cite{Aly:2015:AAQ:2831360.2831361}). We chose the maximum cache size setting based on \cite{Sundaram:2013:SSS:2556549.2556574}. 

\subsection{Discussion of the Performance Results}
We first show the effect of different values of top-k for the Audio and the P53 dataset in Figures \ref{fig:kAudio} and \ref{fig:kP53} respectively. From these figures, it can be seen that k does not have a noticeable effect on the performance of the algorithms. Hence, in further charts, we only consider k=50 due to space limitations and for simplicity purposes. In our experiments, we also observed that the alternative, QALSH\_Cd, always gave the worst result (because of insufficient space to cache the index files). Hence we also omit QALSH\_Cd from the charts. 

In Figure \ref{fig:defaultSettings}, we compare \textit{qwLSH} against all alternatives on the 4 real datasets for the default settings [16MB cache, |Q|=250]. While for Audio, LabelMe, and Mnist, our proposed model always matches with the optimal setting (QA\_Opt), QA\_Opt for P53 returns a slightly lower IO than \textit{qwLSH} (154.66 vs. 157.22). This is because our model returns an Index Cache size setting of 45.41\%. Since we only compare the setting sizes with increments of 10 (i.e. for 40\% and 50\%), our model does not return the most optimal answer. Even with this small drawback, the difference between the IOs is small. For LabelMe and Sift, the Index Cache size setting of 99\% is always the most optimal because the IndexIO cost is always very dominant. Due to this reason (and space limitations), we only present the results for Audio and P53 datasets in our next experiments. 

\noindent \textbf{Effect of Varying Total Cache Size: }Figure \ref{fig:diffCacheSize} shows the effect of different cache sizes on \textit{qwLSH} and its proposed models. It can be seen that our proposed model can adapt to different cache sizes. While \textit{qwLSH} does not always return the most optimal setting, it is still always very close to the optimal setting. For some scenarios (e.g. Audio, 8MB), the optimal answer is close to QA\_Ci, whereas for some scenarios (e.g. P53, 20MB), the optimal answer is closer to QA\_CiCd. In both these scenarios, \textit{qwLSH} is able to adapt and return results closer to the optimal setting. 

\noindent \textbf{Effect of Varying Number of Queries in the Query Workload: }Figure \ref{fig:diffQueries} shows the effect of different number of queries in the query workload. It can be seen from Figure \ref{fig:diffQueries} that the optimal answer for Audio is closer to QA\_Ci, whereas the optimal answer for P53 is closer to QA\_CiCd. \textit{qwLSH} can always adapt and return results closer to the optimal setting.

\begin{figure}
	\centering
	\includegraphics[width=0.8\linewidth]{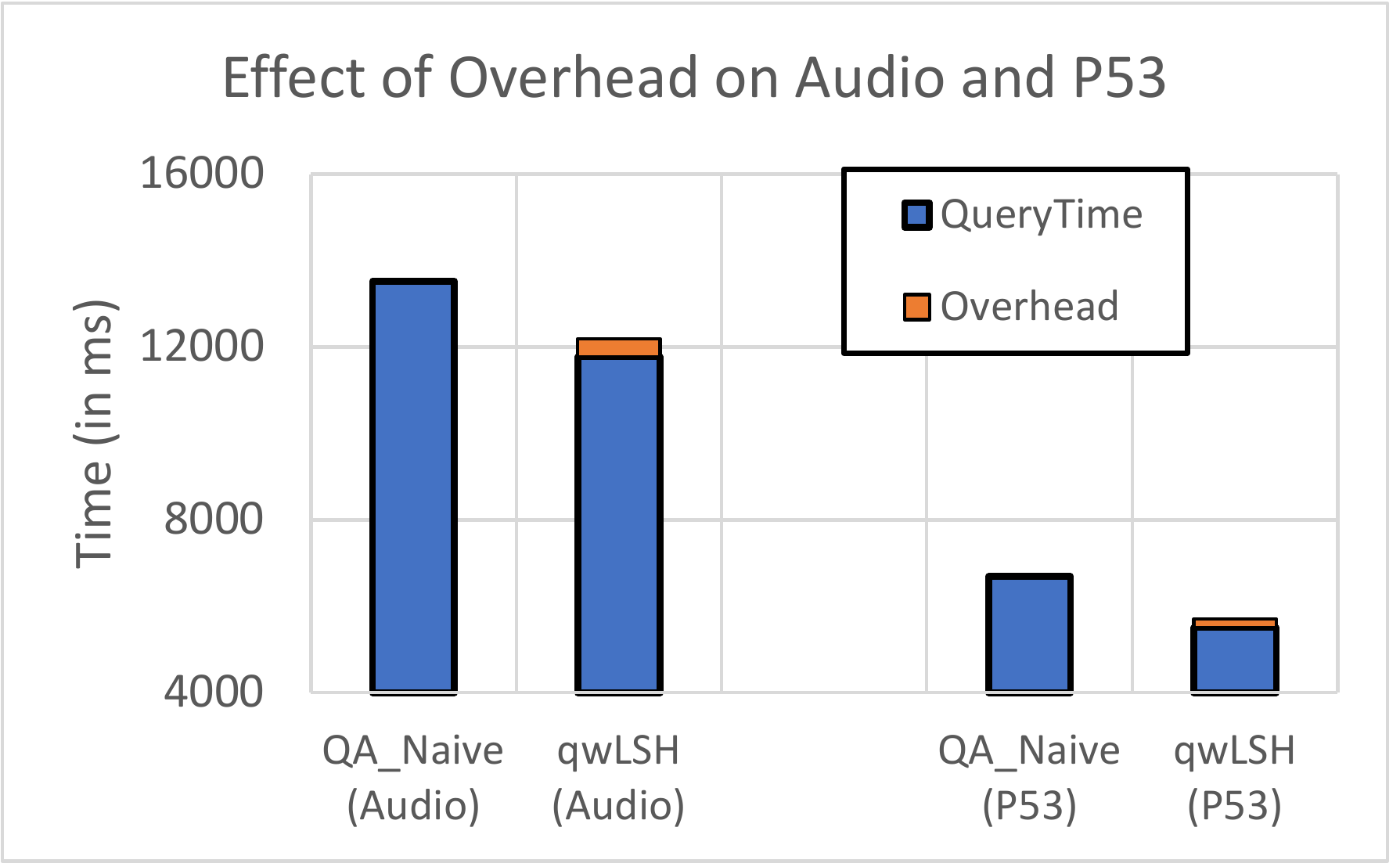}
	\caption{Effect of caching overhead on Time (in ms) }
	\label{fig:time}
\end{figure}

\noindent \textbf{Overhead of qwLSH's Cache Implementation: }Figure \ref{fig:time} shows the negligible overhead of \textit{qwLSH} when compared with QALSH\_Naive for Audio and P53 datasets. Even with the caching overhead, \textit{qwLSH} is still faster than QALSH\_Naive because of the efficient cache utilization and the resultant savings in the total IO. 

\section{Conclusion}
\label{sec:concl}
In this paper, we presented a novel cache-conscious index structure, \textit{qwLSH}, for efficient execution of query workloads in high-dimensional spaces. Traditional LSH-based index structures are not designed to efficiently query workloads in high-dimensional spaces. Based on important observations about the effect of cardinality and dimensionality of a dataset on cache utilization, we intelligently divided a given cache during processing of a query workload by using novel cost models. Experimental analysis over different real datasets under different settings showed the effectiveness of \textit{qwLSH}. 

\newpage

\bibliographystyle{plain}
\bibliography{references}

\end{document}